\documentclass[a4paper,11pt]{article}
\usepackage{jinstpub} 

\usepackage{caption}
\usepackage{subcaption}
\usepackage[table]{xcolor}
\usepackage{siunitx}
\usepackage[section]{placeins}
\newcommand{\emittance}{\pi\, \mathrm{mm}\, \mathrm{mrad}}

\title{Realization and simulations of the new SPES Beam Cooler}

\author[a,1]{A. Ruzzon,\note{Corresponding author.}}
\author[a]{M. Maggiore}
\author[a]{C. Roncolato}
\author[b]{G. Ban}
\author[b]{J.F. Cam}
\author[b]{C. Gautier}
\author[b]{C. Vandamme}
\affiliation[a]{Laboratori Nazionali di Legnaro, Istituto Nazionale di Fisica Nucleare, Viale dell'università 2, Legnaro, Italy}
\affiliation[b]{Laboratoire de Physique Corpuscolaire, 6 Bd Maréchal Juin, Caen, France}

\emailAdd{ruzzon@lnl.infn.it}

\abstract{One of the aim of the new project Selective Production of Exotic Species (SPES) is to produce and select new, neutron reach, isotopes \cite{GPrete_2009,GalataRSI2014,deAngelis_2015,andrighetto2015spes}. The technique adopted to produce these species is the Isotope Selection On Line (ISOL) that postpones the separation of the isotopes after the beam production, using a High-Resolution Mass Spectrometer (HRMS).

In order to allow a good separation of isotopes , the transverse emittance and the energy spread of the beam should have very low values, for this reason the Beam Cooler (BC) is located between the ISOL target, i.e. the beam source, and the HRMS.

In the SPES project, where heavy isotopes are expected (mass number >100), the spectrometer resolution must be preferably higher than $dm/m$=\SI{5E-5}{} and thus the features of the beam at the entrance of the HRMS should be at least  $\varepsilon^n_{95\%}<$8.3E-3 $\emittance$ and $\sigma_E<$\SI{1.5}{eV}.

BC devices are devoted to improve the beam features in terms of emittance and energy spread. A new BC has been designed and realized by the \emph{Laboratoire de Physique Corpuscolaire} (LPC) at Caen, France, for the SPES facility at \emph{Laboratori Nazionali di Legnaro} (LNL), near Padova, Italy.

BCs cool down the beam thanks to a dissipative process in which the thermal energy passes from the beam ions to another medium whose constituent is typically much lighter, Helium gas in our case. This process takes place inside a confinement system that on the one hand to limit the spread of the cooling medium, and on the other to allow the beam to continue along the required trajectory, in the presented device it is a row of radio frequency quadrupoles in an almost closed chamber.

Some specifications of the BC described in this document: the RFQ is \SI{723}{mm} long, its internal radius is \SI{5}{mm}; the RFQ is included in the gas chamber which is \SI{730}{mm}$\times$\SI{280}{mm}$\times$\SI{220}{mm}. Before it there is the injection part composed by, following the beam trajectory, a grounded pipe (\SI{388}{mm} long) and three focusing electrodes. On the opposite side there is the extraction part with two focusing electrodes.

This document presents the main features of the new BC together with the results of a preliminary study where the beam dynamic has been simulated. The analysis embeds also the investigation of the gas distribution inside and outside the BC.

The beam dynamic simulations are based on the Simion code \cite{simionWebsite} while the estimation of the gas distribution is computed with MolFlow+ \cite{MolflowKersevan2019}.

Simulations show that accurately setting the BC leads to a large improvement of the emittance while the energy spread still needs to be improved. Limiting the gas pressure in the acceleration zone seems to allow the required final boost.}


\arxivnumber{} 

\begin{document}
	\maketitle
	\flushbottom
	
\section{Introduction: operation principles of a RFQ beam cooler} \label{sec:intro}

A BC is devoted to improve the transversal and longitudinal emittance of the beam transferring its thermal energy to another medium. When the other medium is a gas, like in the device presented here, the transfer is dissipative occurring through interactions between ions and gas and therefore the beam will not retain the memory of its initial characteristics.

In order to allow the best condition for the energy transfer the gas must be kept at relatively high pressure and the beam has to be slowed down to a few tens of eV. In order to ensure sufficiently high pressure and good gas confinement a chamber equipped with the beam inlet and outlet, which consist of two holes kept as small as possible, is required. At the same time the beam slowdown is obtained maintaining that chamber at a voltage close to the beam energy. Typically the gas pressure is few pascals and the kinetic energy left to the beam at the beginning of the process is about one or two hundreds of eV.

Since the ions-gas interaction will induce to diffusion and the loss of the beam, also a confinement system for charged particles is required. The most common of these systems, also installed in the presented device, is the linear Paul's trap, which consists of a long quadrupolar device with a radio frequency voltage that confine the beam along its axis. The transverse motion of ions in such a trap is, to first order, harmonic.

So far, along the axis of the quadrupole, ions move freely and thus they can return back to the inlet, for this reason the confining quadrupole is longitudinally divided in several parts with a small and constant potential gradient that imposes the ions to move towards the outlet.

Inside a Paul's trap the electric potential is given by:
$$
\Phi = \frac{1}{2} \frac{\Phi_0}{r_0^2} \left(x^2 - y^2\right)
$$
Where $\Phi_0$ is the polarization potential and $r_0$ is half of the distance between the opposite poles of the quadrupole. In the considered quadrupole the potential is periodic: $\Phi_0 = U_{RF} \cos(\omega_{RF} t)$ with $U_{RF}$ and $\omega_{RF}$ the amplitude and the pulsation respectively.

From the above definition, the motion of a charged particle is described by these equations of motion:
\begin{eqnarray}	
	\ddot x + \frac{2 e}{m r_0^2} \Phi_0 x & = & 0 \nonumber \\
	\ddot x - \frac{2 e}{m r_0^2} \Phi_0 y & = & 0 \nonumber \\
	\ddot z & = & 0 \nonumber
\end{eqnarray}
In these last equations $m$ is the particle mass.

The last equations can be written in a more general form using the Mathieu's coordinates:
\begin{eqnarray}
	\tau & = & \frac{\omega_{RF}}{2} t \nonumber \\
	q & = & q_u = q_x = - q_y = \frac{4 e U_{RF}}{m \omega_{RF}^2 r_0^2} . \nonumber
\end{eqnarray}

The results represent a particular form of the Mathieu's equation that has solutions of this form:
\begin{eqnarray}
	u(\tau) = \alpha' e^{\mu \tau} \sum_{n=-\infty}^{\infty} C_{2n} e^{2ni\tau} + \alpha'' e^{-\mu \tau} \sum_{n=-\infty}^{\infty} C_{2n} e ^{-2ni\tau} , \nonumber
\end{eqnarray} 
in which $\alpha'$ and $\alpha''$ depend on the initial conditions of $u$, $\dot u$ and $\tau$, while $ C_{2n}$ and $\mu$ depend on $q$ \cite{mcl51}.

With appropriate simplifications, solutions can be rewritten as a linear combination of the following functions.
\begin{eqnarray}
	u_1(\tau) = \sum_{n=-\infty}^{\infty} C_{2n} \cos( (2n + \beta) (\tau - \tau_0) ) \nonumber \\
	u_2(\tau) = \sum_{n=-\infty}^{\infty} C_{2n} \sin( (2n + \beta) (\tau - \tau_0) ) \nonumber
\end{eqnarray}

In this way, the trajectory results mainly as a superposition of two oscillations with different amplitudes and frequency, depending on the initial conditions and on the quadrupole settings.

A stability analysis reduces the available $q$ to some intervals among which the most interesting is between $ [ 0, 0.91 ] $. More details about the theoretical dynamic of an ion in a RF quadrupole can be found in \cite{daw95}, \cite{Ars64} or \cite{mcl51}.

The above theoretical analysis defines a constrain about the bias settings of the quadrupole and tells us also that ions have a coherent motion induced by the RF field. A further analysis of such a motion in the presence of a gas suggests the existence of an undesired effect called \emph{RF heating} \cite{kimThesis}: as a result of the gas interaction an ion, which is moving coherently with the RF field, changes abruptly its energy state, that typically increases. The heating can be zeroed choosing values of $q$ as low as the confinement allows.

From the synthetic analysis presented above, the most important parameters that determine the performance of a BC are: its overall dimension, the quadrupole setting, the gas distribution and density and the way the beam is injected and extracted from the device.

The present work details the physical construction of the new BC for SPES in \autoref{sec:the_cooler}, while \autoref{sec:gas_profile} shows how the filling gas is expected to behave inside and outside the main chamber, \autoref{sec:simion_simulation_intro} and \autoref{sec:simion_simulation} report the results of the beam dynamic simulations that are discussed in \autoref{sec:discussion} and finally \autoref{sec:conclusions} drowns some conclusions.

\section{The new Beam Cooler for SPES} \label{sec:the_cooler}

In this section some specifications on the BC of the SPES project are reported.

For the sake of synthesis we can start from the main component of this device which is the RFQ: it is \SI{723}{mm} long and has an internal radius of \SI{5}{mm}, it is composed by 18 sectors, which in turn are composed by 4 rods with a radius of \SI{5.7}{mm} each. The four rods are supported by four stainless steel bars (\SI{20}{mm}$\times$\SI{10}{mm}) and each rod is supported by two PEEK insulators fixed in the bar. Thanks to those insulators one can adjust the position of the rods.

The quadrupolar electric field for the beam confinement is produced by applying two sinusoidal voltages in opposite phase on the four segmented rods of the RFQ. The internal radius of the quadrupole was defined by simulations and frozen by construction while the amplitude and the frequency of the sinusoidal voltages are adjustable parameters allowing to move in the stability diagram of the RFQ cooler (\cite{Boussaid_2014}). The RFQ is powered by a RF generator (a KEYSIGHT 33509B) and a RF power amplifier (a Rohde \& Schwarz BBA100-A500 with an output power up to \SI{500}{W}).

The guidance of ions from the input to the output of the gas chamber is ensured by a longitudinal electric field which is produced by a ramp of DC voltages applied to each rod segment.

The RFQ is included in the gas chamber (\SI{730}{mm} long \SI{280}{mm} high and \SI{220}{mm} large) which constitutes the central structural component of the BC. The entire BC is located on an insulated platform with a High Voltage (HV) insulator transformer by TRANSFO.Industrie able to reach 50 kV with a very small ripple. In any case, the platform is equipped with a specific dispersion circuit for the mitigation of unwanted ripples that can be controlled remotely.

Before the gas chamber there is the injection part which is \SI{470}{mm} long and composed by, following the beam trajectory, a grounded pipe (\SI{388}{mm} long) and three focusing electrodes. On the opposite side there is the extraction part, symmetrical to the injection one, except for the focusing electrodes that here are two and differently designed. All the injection and extraction lenses can be independently polarized from 0 to \SI{3}{kV}, considering the HV platform as the reference.

\begin{figure}[!hbt]
	\centering
	\includegraphics*[width=.9\columnwidth]{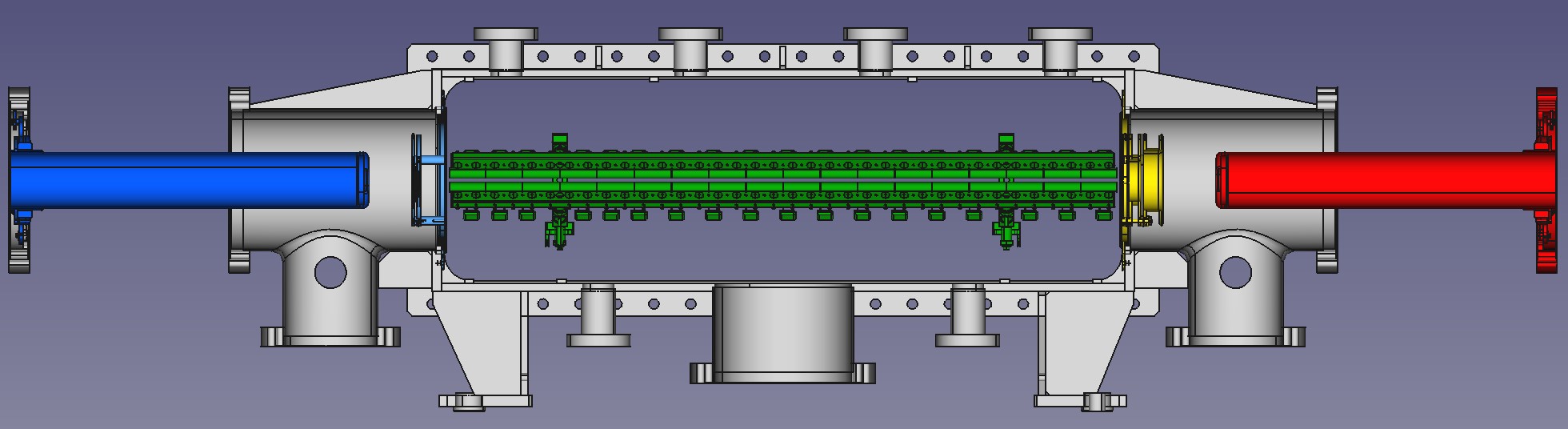}
	\caption{A sketch of the BC. The beam proceeds from the left to the right: the injection ground electrode (red), the injection lenses and iris (yellow), the RFQ composed by 18 sectors, the extraction iris and its lenses (light blue), the extraction ground electrode (blue).}
	\label{fig:BC_scheme}
\end{figure}

A sketch is in \autoref{fig:BC_scheme}.

The cooling effect is ensured by the injection of helium into the quadrupole chamber whose flow is controlled by a Mass Flow Controller (MFC): it is a BROOKS SLA5850, which is able to control the gas flow from 0 to \SI{150}{slpm} with a <1\% error.

Each compartment (injection, gas chamber and extraction) is equipped with a turbo pump, but during the cooling activity the central turbo pump is closed in order to minimize the gas flowing through the gas waste.

\begin{figure}[!hbt]
	\centering
	\includegraphics*[width=.7\columnwidth]{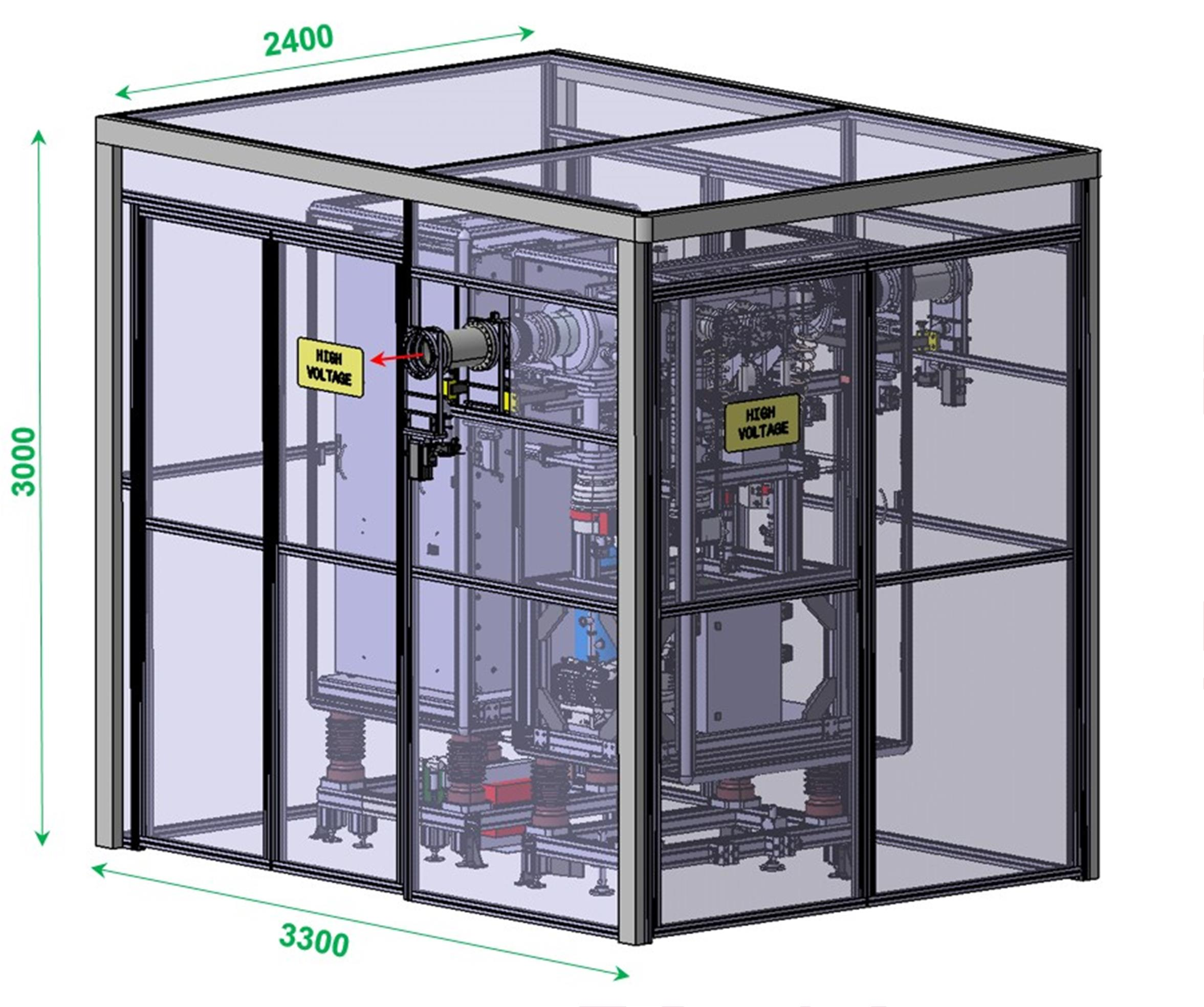}
	\caption{A render of the BC inside its safety cage.}
	\label{fig:BC_assembly}
\end{figure}

Since the BC is located on an insulated platform the entire device is inserted in a protection cage \SI{2.4}{m} long, \SI{3.3}{m} large and \SI{3}{m} high (the beam height in SPES is \SI{2.32}{m}). \autoref{fig:BC_assembly} shows the BC inside its safety cage.

The remote control of the BC is based on the EPICS standard with a client-server configuration: the operator will supervise the devices through a graphic user interface in a client PC, which can run also scripts for automated operations, while all the BC variables are managed in the server \cite{Marcato:2022kbb}. The server is then connected to the devices for the RF setting and to the main PLC, a Modicon Series from Schneider (splitted in ground and HV component), that manages mainly the vacuum devices and the interlocks.

The safety system, namely the management of the access to the HV area, runs on a independent PLC, manufactured by Pilz.

The BC assembly was completed in 2021 at the Labortoire de Physique Corpuscolaire (LPC) at Caen, France and the functional tests began later in the same year. To complete the commissioning performance tests with four different beams are on going and their end is planned during the first half of 2024 at LNL. 

\section{Gas distribution study}\label{sec:gas_profile}

This section explains how the pressure profiles in the iris regions are computed.

In the first part we overview the actual BC state in order to implement a plausible gas density in the beam dynamic simulations. In the second part small geometrical modifications are introduced in the BC design in order to limit the gas leaks which increases the beam emittance and finally to improve the BC performances.

\subsection{Gas profile with the native BC design} \label{subsec:native_gas_profile}

As described in the previous paragraphs, the BC is composed by a box, namely the HV chamber, in which the gas is injected and which design is conceived to let the gas leaks only through the injection and extraction holes. Its big transverse dimensions should ensure the homogeneity of the pressure along the longitudinal axis while the small sizes of the holes should imply in a sharp drop of the pressure outside the chamber.

One of the most important results obtained in the next section is that the beam dynamic simulations resulted very responsive to the gas distribution outside the chamber, both in the injection and extraction area, and thus it must be carefully estimated. Such an estimation is obtained simulating the gas leakage with the Monte Carlo code MolFlow+ \cite{MolflowKersevan2019}. It allows to calculate the pressure in a complex geometry whenever the gas behavior can be considered molecular.

\begin{figure}[!tb]
	\centering
		\includegraphics[width=0.7\textwidth]{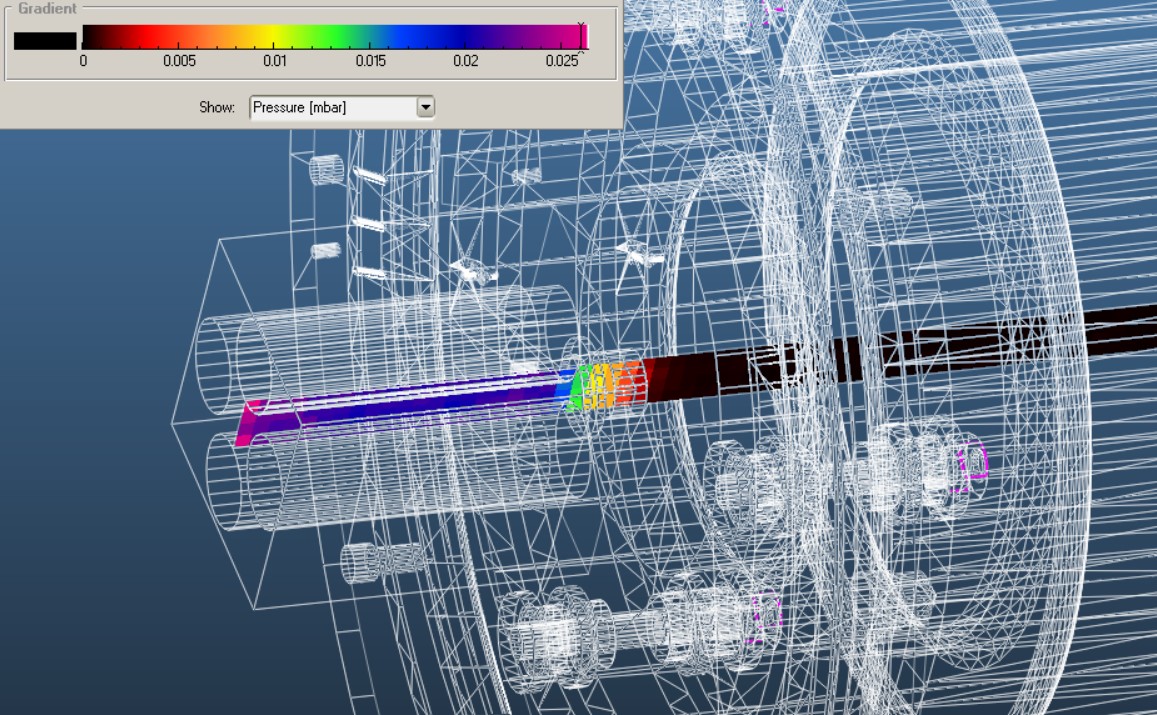}
	\caption{The injection iris as it appears in the MolFlow+ GUI. The colored vertical rectangle is the virtual surface recording the gas pressure.}
	\label{fig:Molflow_1}
\end{figure}

In the present case, the entire chamber with the expected pressure does not fulfill the molecular regime requirement (inner typical dimension $D \sim 10$ cm, expected pressure $P \sim 3$ Pa and thus Knudsen number $K \sim 0.03$), but reducing the considered region close enough to the irises implicates smaller dimensions and pressures and thus the gas behavior will be molecular. This is why the following simulation is going to consider, inside the chamber, just the quadrupoles closer to the irises and a small region around them. Vice versa the volumes outside the chamber, where the pressure is lower than 0.1 Pa, are entirely considered. \autoref{fig:Molflow_1} shows these details as presented by the Molflow+ GUI.

\begin{figure}[!tbh]
		\centering
		\includegraphics[width=0.45\textwidth]{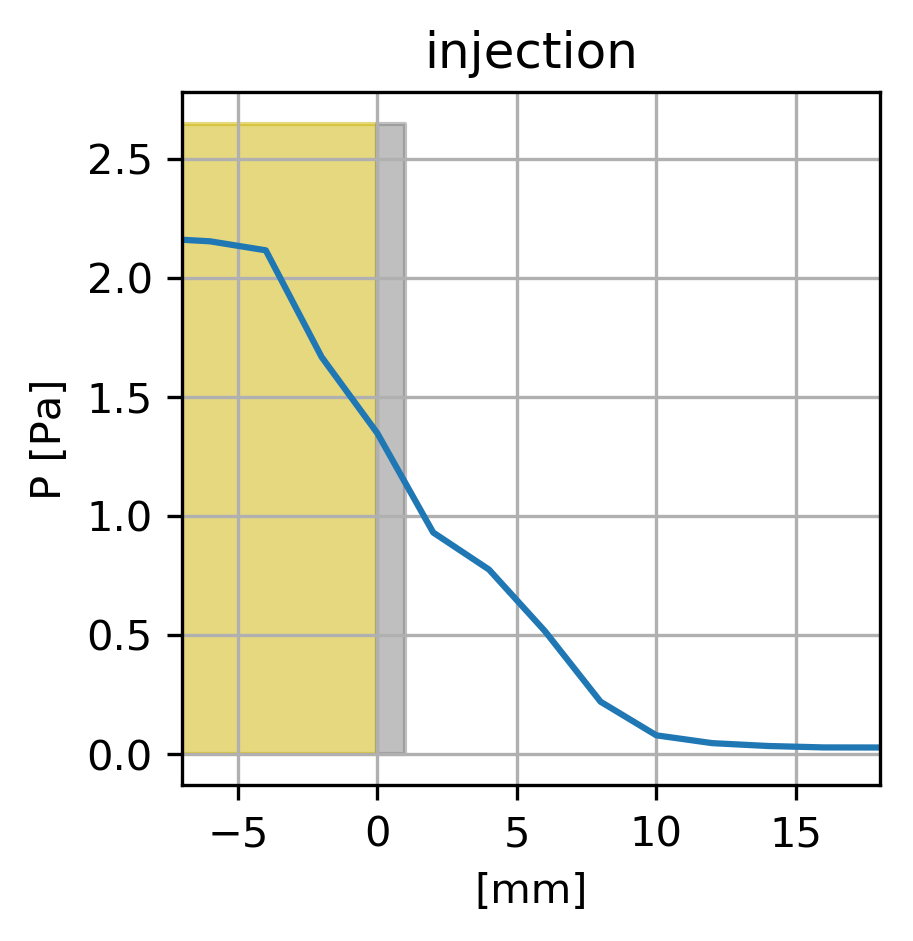}
		\includegraphics[width=0.45\textwidth]{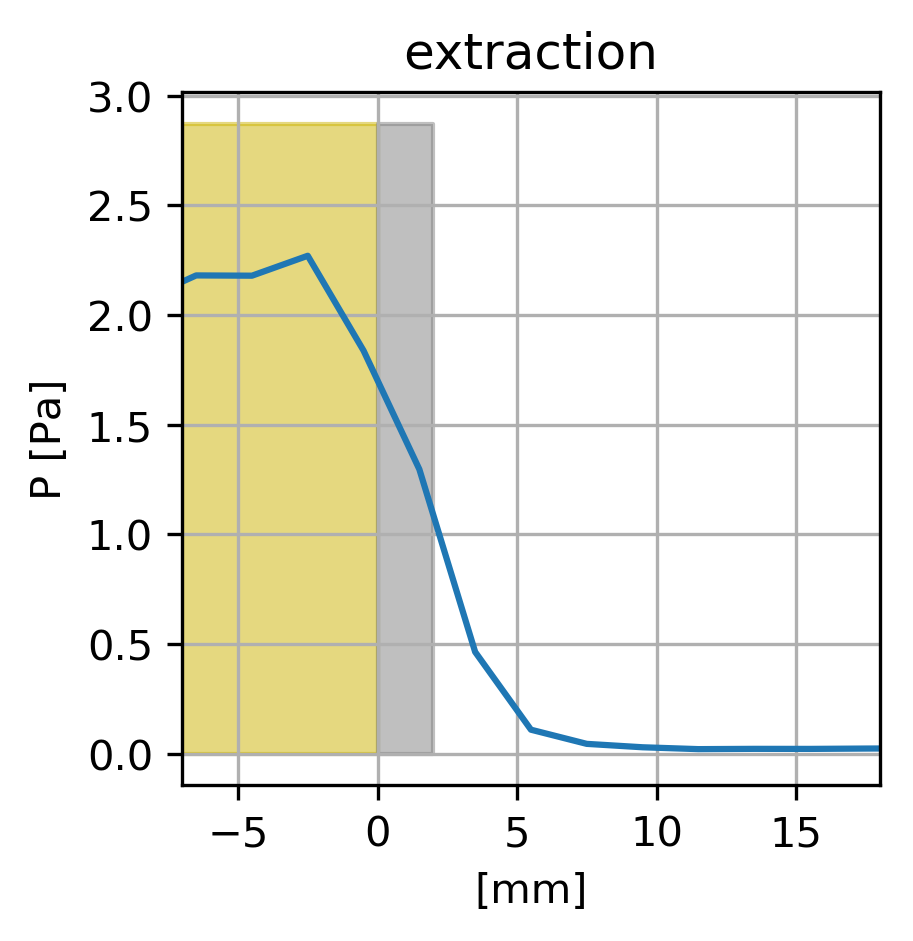}
	\caption{The pressure profiles close to the two irises: injection on the left and extraction on the right. The \textit{x} axis shows the position in millimeters centered in the iris, the yellow rectangle is the volume inside the HV chamber and that gray is the iris thickness.}
	\label{fig:P_profile_1}
\end{figure}

Such a simulation gives the pressure profiles reported in \autoref{fig:P_profile_1}. Both the curves present similar features: they start to decrease few millimeters before the iris with an almost linear pace that reaches a negligible value after some millimeters outside the iris, but they differ in the distances at which this happens. In the injection iris the drop starts at -4 mm and ends at 10 mm, while in the extraction iris, due to the smaller diameter (6 mm here against 8 mm before), it starts at -2 mm and ends at 5 mm.

Such profiles are inserted in the Simion code in order to get a reliable beam simulation.

\subsection{Gas profile with a modified extraction iris} \label{subsec:modified_gas_profile}

The ideal BC has gas perfectly confined in the HV chamber in order to avoid collisions between ions and gas atoms in those zones without a confinement system or where an important accelerating field takes place, i.e. where they are likely to increase the beam energy spread.

For the sake of clarity we anticipate here one of the results stated in the next section, i.e. the pressure leakage is the most effective parameter in the energy spread control of the cooled beam, and thus all the ways to control the gas diffusion represent an improvement. The pursued effect is to avoid the ion-gas interaction during the acceleration of the beam, since every impact will spread the ions energy. Here we present two simple but effective solutions that can be easily implemented in the future.

\begin{figure}[!tbh]
	\centering
		\includegraphics[width=0.45\textwidth]{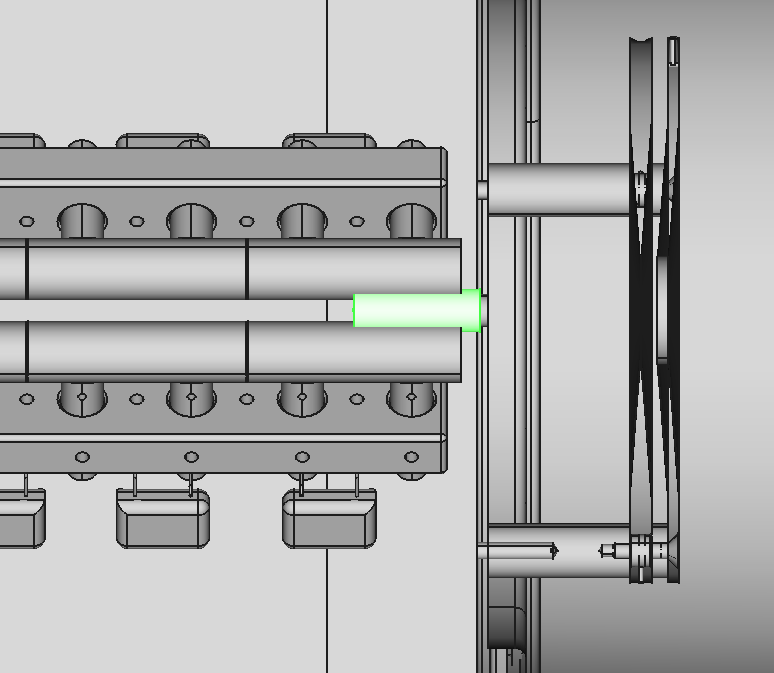}
	\caption{The pipe added to the extraction iris. The picture represents the mechanical sketch. Here the beam travels left to right finding in sequence: the last two quadrupole rods with their support, the pipe (in green), the septum separating the inside from the outside of the chamber and the two electrodes of the extraction lens.}
	\label{fig:pipe}
\end{figure}

The first is a small pipe that can be positioned in the last quadrupole and that reduces the conductance of the gas from inside to outside the chamber and the emittance. Such a pipe, with a length of 23 mm, is shown in \autoref{fig:pipe}. The second solution is the reduction of the extraction hole diameter from 6 mm to 3 mm.

In order to account the increased geometrical complexity inside the HV chamber these last two gas simulations rely on a simplified model which includes the entire volume of the chamber. Such an approach has the advantage of allowing the verification of the pressure trend along the entire quadrupoles axis and the reliability of a pressure measurement taken at the side of the chamber.

\begin{figure}[!tbh]
	\centering
		\includegraphics[width=0.45\textwidth]{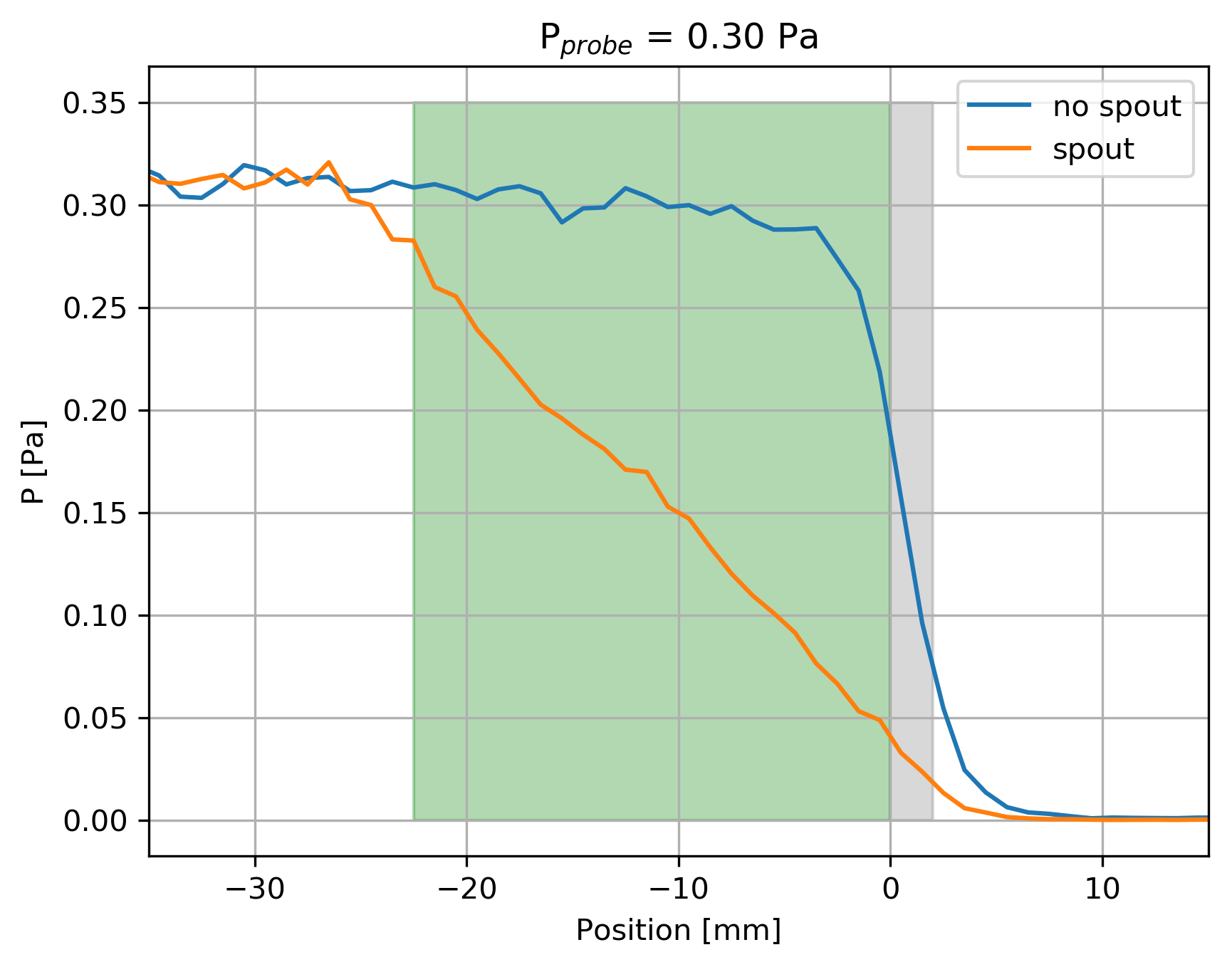}
		\qquad
		\includegraphics[width=0.45\textwidth]{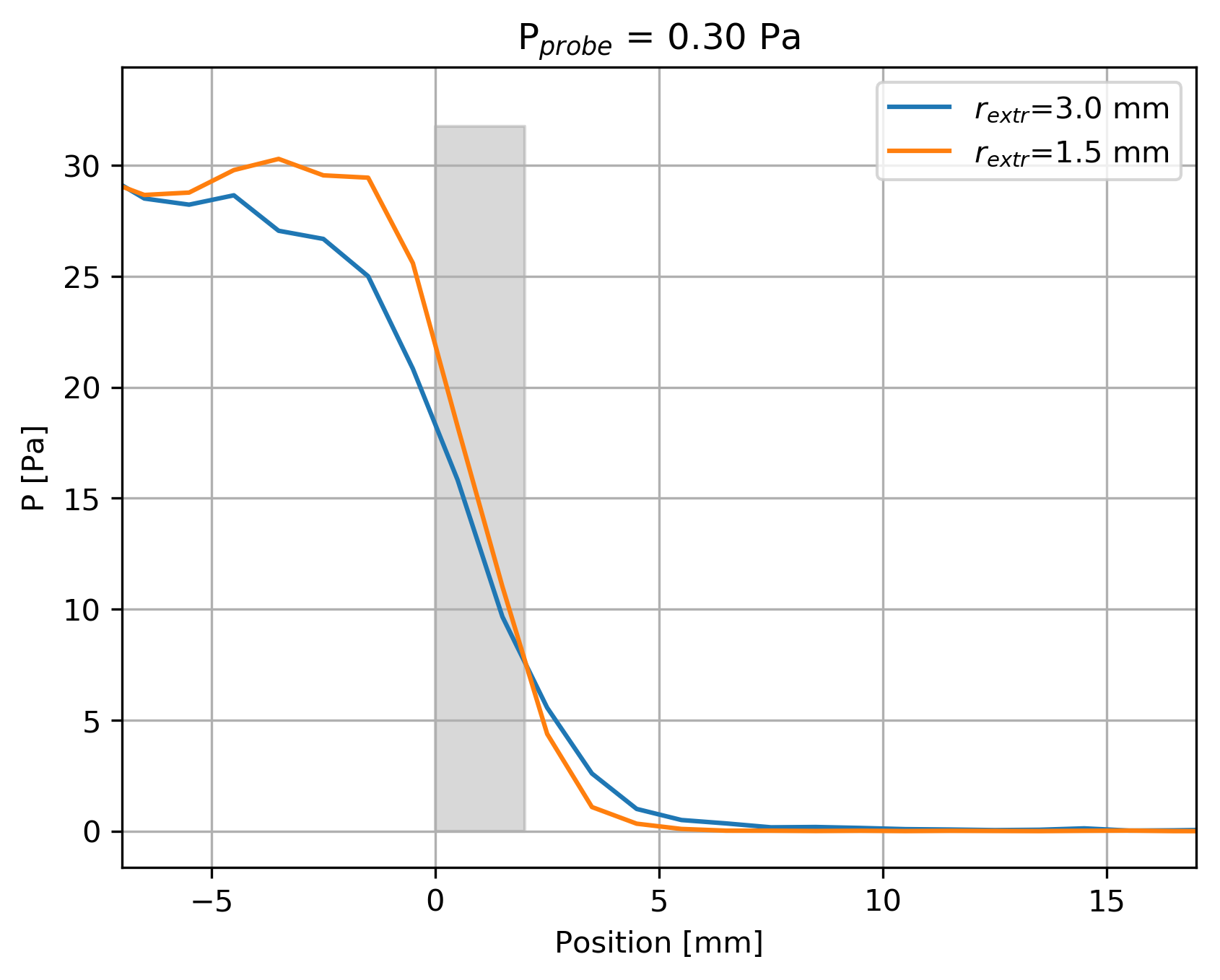}
	\caption{The left picture represents the comparison of the pressure profile with and without the additional pipe (orange and light blue line respectively). The pipe interval is underlined by the green area.}
	\label{fig:pipe_reduction_profile}
\end{figure}

The new pressure profiles, compared to that obtained with the old configuration, are shown in \autoref{fig:pipe_reduction_profile}.

In the pipe case the pressure gradient is lower and it starts to drop more internally, this allows a lower gas density along the acceleration trajectory, that starts just after the last quadrupole.

The reduced hole diameter (from 6 to 3 mm) implies a sharper pressure gradient which results in a slightly higher pressure between the last quadrupole and the iris and a lower gas density in the outer part of the chamber, that is where the strongest acceleration takes place.

\section{Beam dynamic simulations} \label{sec:simion_simulation_intro}

The beam dynamic has been simulated with the Simion code edition 8.1 \cite{simionWebsite}. It can compute the electric fields of a configuration of electrodes and the trajectories of charged particles starting from their initial conditions. Moreover it can include an user program file to account electrode properties, like the potential time dependency, the gas distribution and the ion-gas collisions. The latter are estimated with the hard-sphere collision model, a library embedded in Simion, where the expected frequency of collisions, measured as a distance (the mean-free-path), is predicted by the kinetic theory of gases as a function of the known pressure, temperature, and collisional cross sections of colliding particles \cite{simionWebsiteCollisions}.

\begin{figure}[!t]
	\centering
		\includegraphics[width=0.8\textwidth]{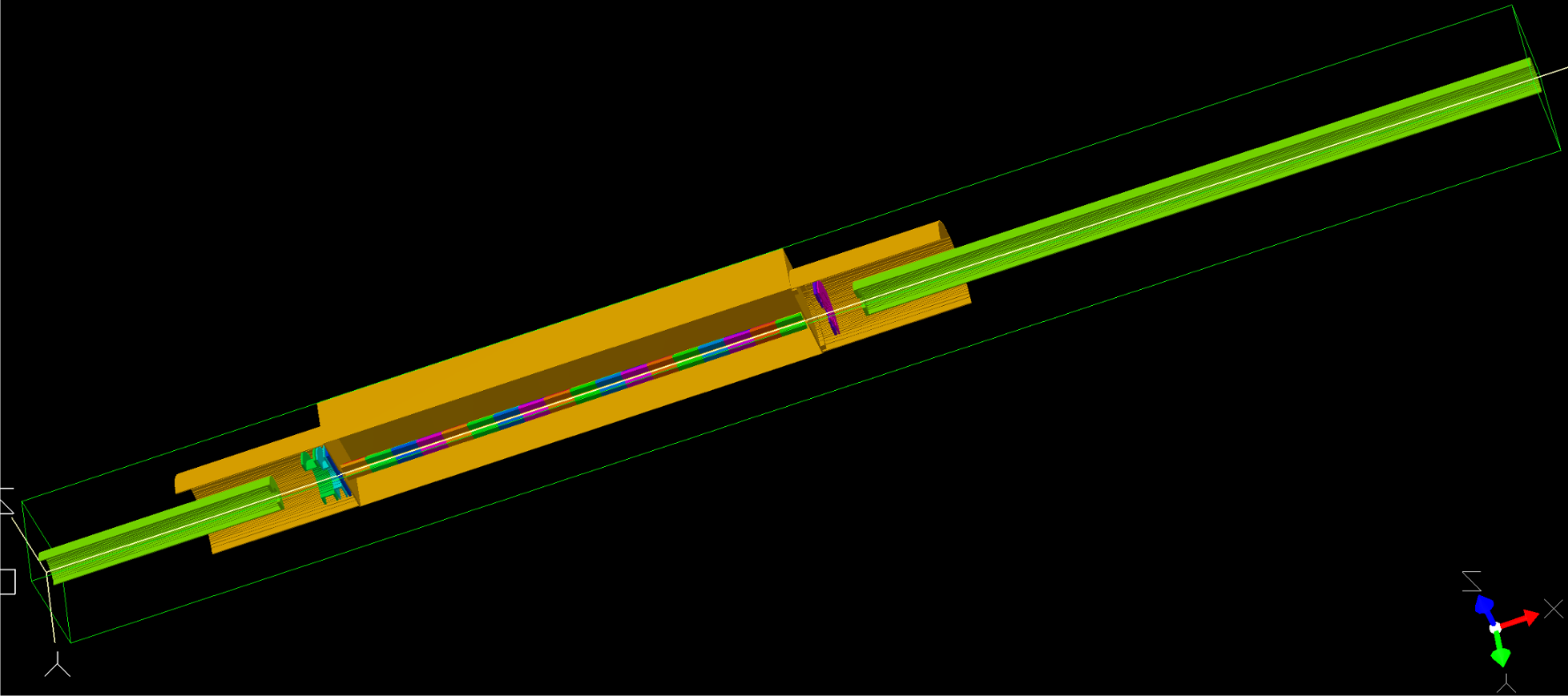}
	\caption{The workbench used to simulate the beam dynamic in Simion. Colors distinguish the references of electrodes.}
	\label{fig:workbench}
\end{figure}

The first step to simulate the BC is to define the \emph{workbench} designing the electrodes configuration. It is a unique volume with a dimension of 2320x80x80 pixels, with a resolution of 1 mm, in which there are the two ground electrodes, the HV chamber, the 18 quadrupole rods and the electrodes composing the injection and extraction lenses. using a single workbench avoids edge errors. The workbench is shown in \autoref{fig:workbench}.

\section{Simulation results} \label{sec:simion_simulation}

In the following results are divided into those for the injection outcome, focused on transmission performance, and those for rest of the BC, focused on beam quality. Two species are simulated: $^{133}$Cs$^+$ and $^{39}$K$^+$.

\subsection{Injection performances}

The most important outcome is transmission. It is computed dividing the number of the particles surpassing the middle of the first quadrupole by their initial number.

Excluding pressure there are two settings that can be changed to improve the injection performance: the HV chamber potential, that imposes the residual kinetic energy of the beam, and the lens setting, which includes three electrodes potentials.

\begin{figure}[!t]
	\centering
	\begin{subfigure}[t]{0.5\textwidth}
		\centering
		\includegraphics[width=0.95\textwidth]{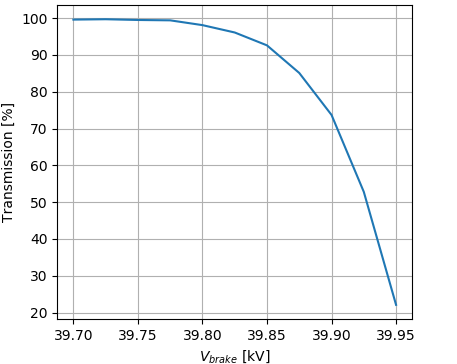}
		\caption{}
		\label{fig:Cs_transmission_brake_potential}
	\end{subfigure}%
	\begin{subfigure}[t]{0.5\textwidth}
		\centering
		\includegraphics[width=0.95\textwidth]{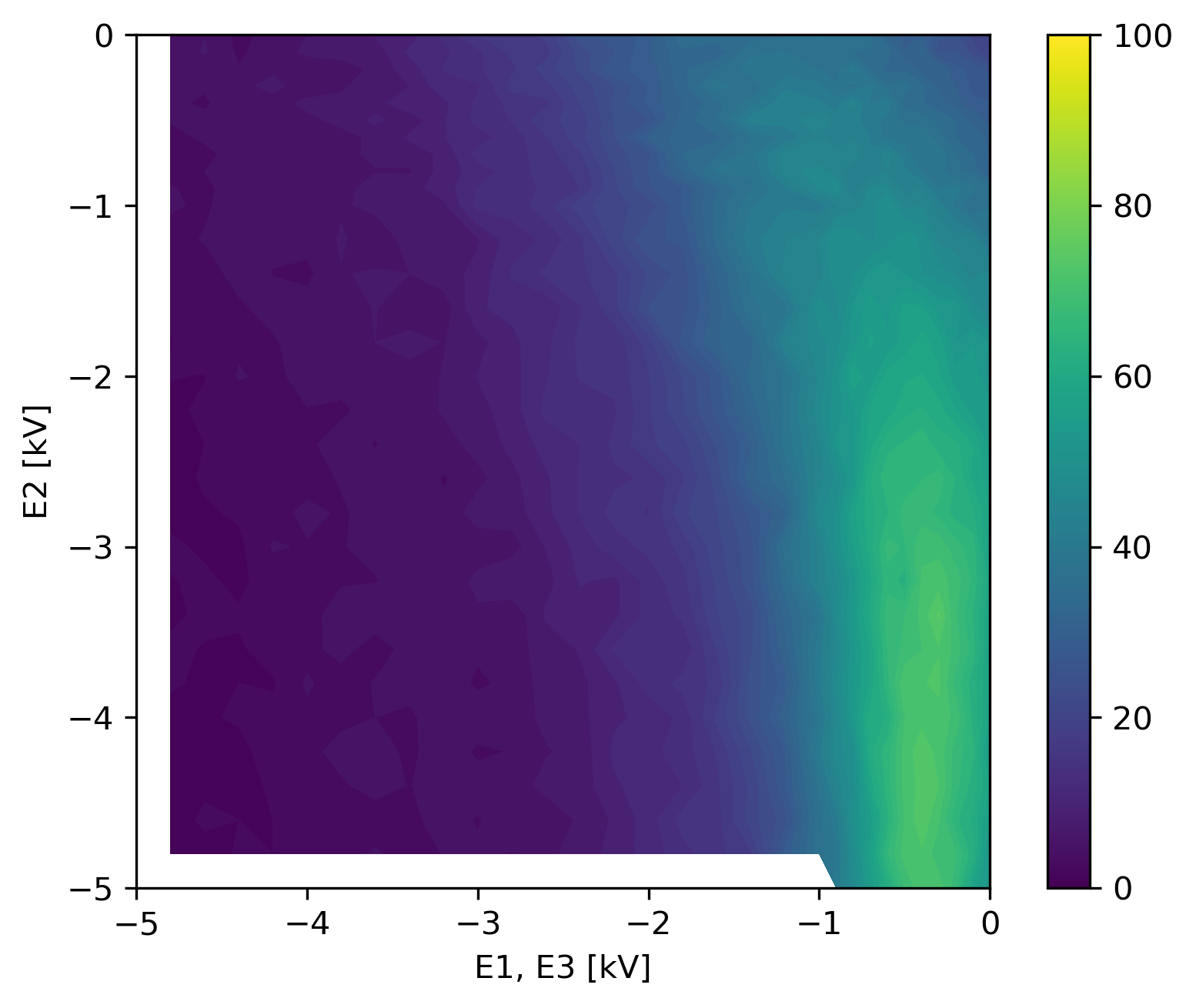}
		\caption{}
		\label{fig:Cs_transmission_injection_lens}
	\end{subfigure}%
	\caption{Injection performance for a beam of caesium. The left picture: the injection transmission (\emph{y} axis) as a function of the HV potential (the \emph{x} axis). In the right: the transmission as a function of the three electrode potentials of the injection lens.}
	\label{fig:Cs_injection_transmission}
\end{figure}

\autoref{fig:Cs_transmission_brake_potential} shows the transmission effect of the HV chamber potential for a beam of $^{133}$Cs$^+$ with initial kinetic energy $E_k = 40\cdot10^3 \pm 0.04$ eV and starting emittance $\varepsilon_{rms} = 10 \emittance$ (this last is the expected value for the ISOL source in the SPES case \cite{manzolaro_RSI_2014}). In this case the gas inside the HV chamber is perfectly confined and has a pressure of 2 Pa, the three electrodes of the lens are set to $V_{E1}=$-1.2 kV, $V_{E2}=$-2.2 kV and $V_{E3}=$-1.2 kV, these potentials are referred to the HV chamber.

The plot reports a transmission close to 100 \% for low stopping potentials, namely with a sufficient kinetic energy left to the ions, vice versa if the chamber voltage gets close to the beam energy, for instance above 39.8 kV, the transmission drops to zero quickly.

The \autoref{fig:Cs_transmission_injection_lens} shows the transmission as a function of the injection lens setting. Its first and third electrode have the same potential and it can be found in the \emph{x} axis, while the second electrode potential is in the \emph{y} axis. The potentials are referred to the HV chamber. The starting beam is equal to the previous case, the chosen stopping potential is 39.74 kV and the pressure is still 2 Pa, but in this case the gas is no longer perfectly confined but has the profile calculated in \autoref{subsec:native_gas_profile}. With the gas extending outside the HV chamber the best performance fell to 75 \% (previous analysis reported a 100 \% transmission for this chamber potential) with a sweet spot at -0.3 V, -3.4 V, -0.3 V for the three electrodes.

\begin{figure}[!t]
	\centering
	\begin{subfigure}[t]{0.5\textwidth}
		\centering
		\includegraphics[height=5cm]{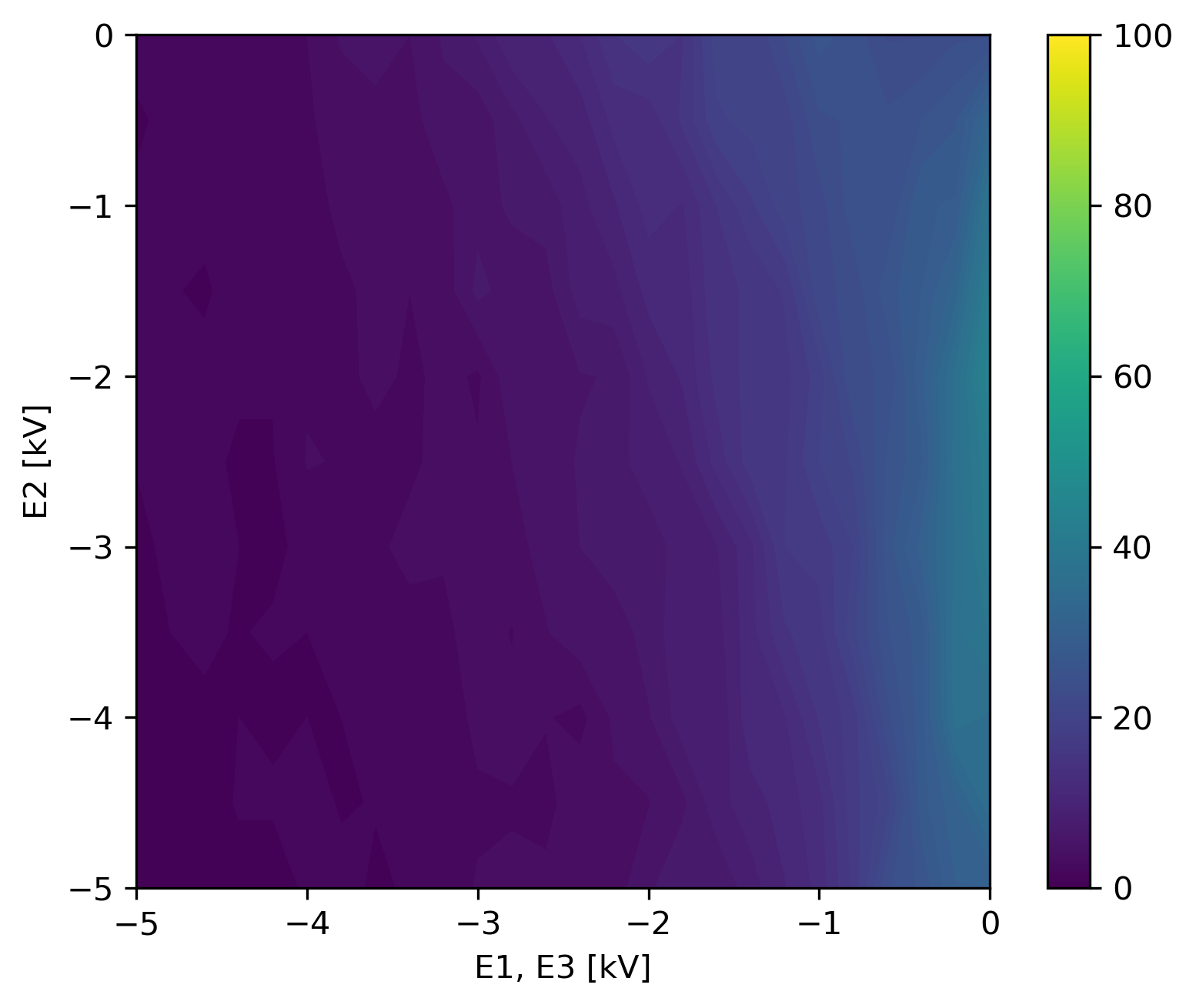}
		\caption{}
		\label{fig:K_transmission_brake_potential}
	\end{subfigure}%
	\begin{subfigure}[t]{0.5\textwidth}
		\centering
		\includegraphics[height=5cm]{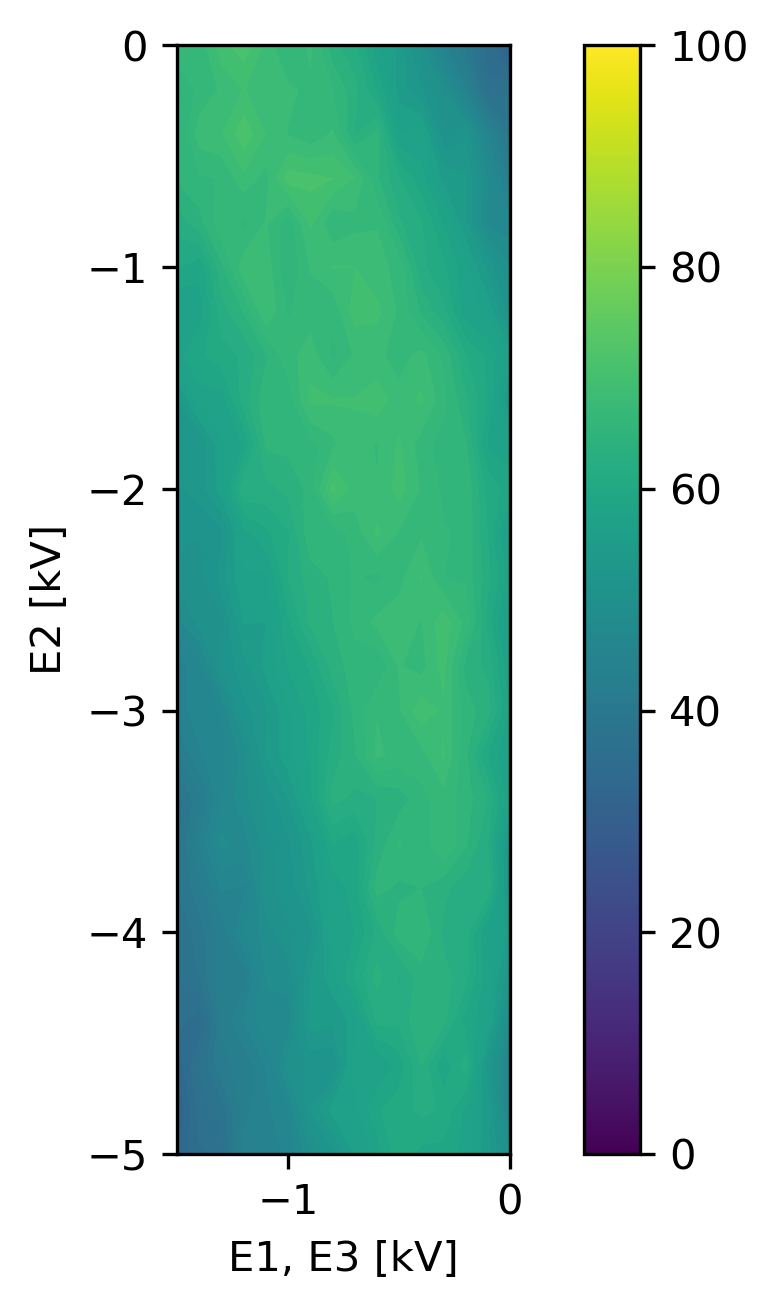}
		\caption{}
		\label{fig:K_transmission_injection_lens}
	\end{subfigure}%
	\caption{Injection performance for a beam of Potassium. The left picture shows the beam transmission varying the injection lens setting: the potentials of the first and the third electrode are equals and are in the \emph{x} axis while the second electrode is in the \emph{y} axis. In this case the pressure is 1 Pa and all the potentials are referred to the HV chamber. In the right there is the same contour but for a pressure of 0.25 Pa.}
	\label{fig:K_injection_transmission}
\end{figure}

Similarly to the \autoref{fig:Cs_transmission_injection_lens}, \autoref{fig:K_injection_transmission} shows the transmission as a function of the injection lens setting, but this time for a beam of $^{39}$K$^+$. The two contours differ for the pressure inside the HV chamber: the left picture is at 1 Pa, the right one is at 0.25 Pa, in both cases the gas is not ideally confined. As in the previous case the transmission can not reach 100 \% but it improves a lot lowering the pressure from 1 Pa to 0.25 Pa, passing from 45 \% to 75 \% respectively. At the same time the sweet spot moves off the axis defined by the $V_{E1}$ = $V_{E3}$ = 0 V to the spot $V_{E1}$ = $V_{E3}$ = -0.5 kV and $V_{E2}$ = -2.0 kV.

\subsection{Complete BC simulations}

In this subsection the simulation results are taken 650 mm after the real end of the BC, which is a reasonable distance to place the first optical element, this is why the simulation workbench is 2320 mm long instead of the BC length.

The parameters varied in these simulations are: quadrupoles tension and frequency, extraction electrodes potential and gas pressure, while the injection elements are kept at the best setup. In the first part of this investigation these parameters are systematically modified with the aim of finding the best compromise between transmission and emittance, in the second part the purpose is to get the best energy spread.

Hereafter the transmission term includes also the losses due to injection.

\begin{figure}[!t]
	\centering
	\begin{subfigure}[t]{0.5\textwidth}
		\centering
		\includegraphics[height=5cm]{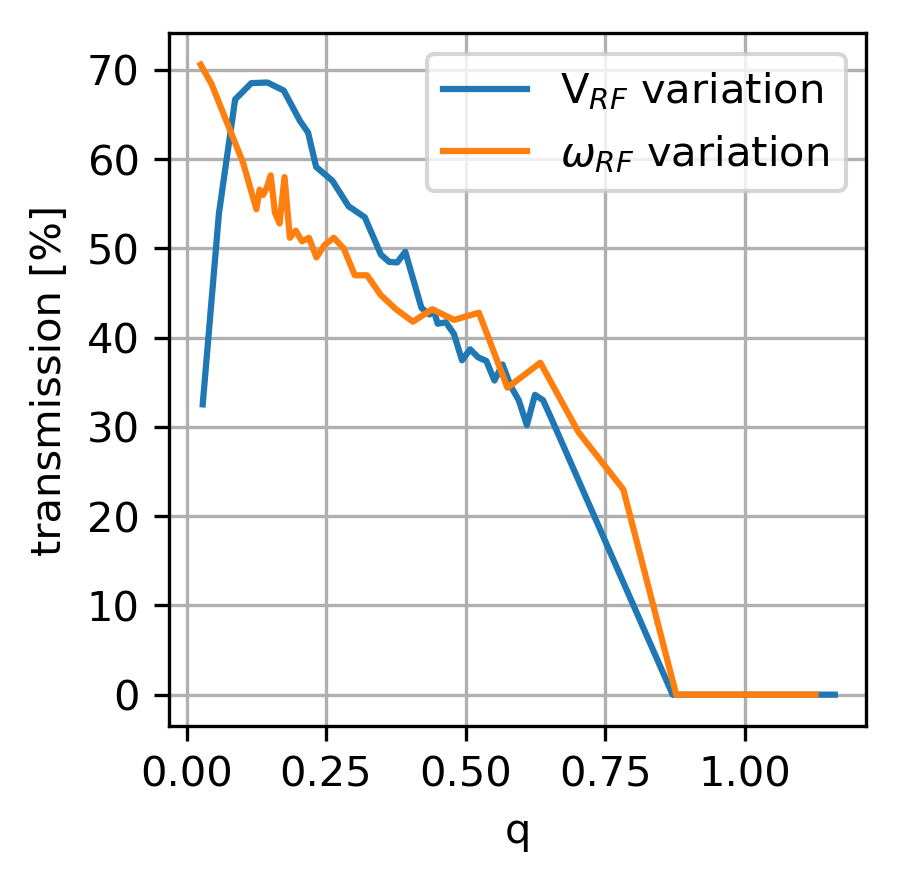}
		\caption{}
		\label{fig:Cs_transmission_q}
	\end{subfigure}%
	\begin{subfigure}[t]{0.5\textwidth}
		\centering
		\includegraphics[height=5cm]{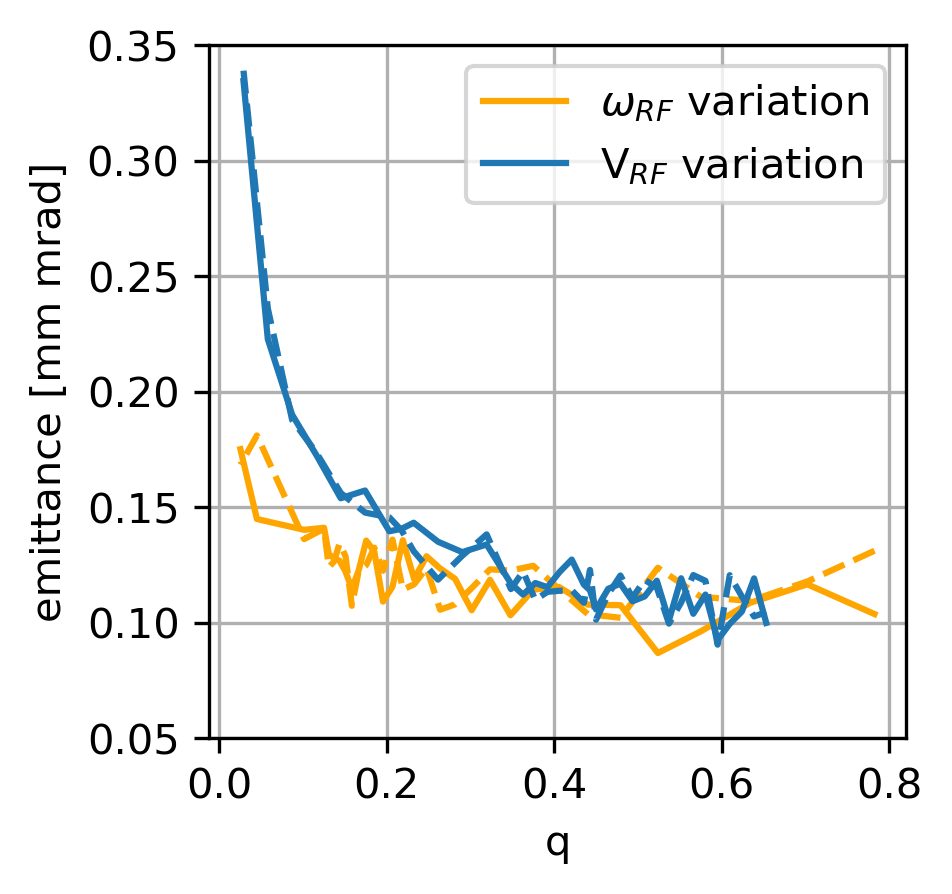}
		\caption{}
		\label{fig:Cs_emit_q}
	\end{subfigure}
	\caption{Effect of the variation of the Mathieu's parameter \emph{q} in the transmission (left) and in the emittance (right) for a beam of $^{133}$Cs$^+$. In the blue curves, \emph{q} is varied changing the polarization potential of the quadrupole, while in the orange curves the polarization frequency changes.}
	\label{fig:Cs_q}
\end{figure}

The quadrupoles settings, i.e. bias voltage and frequency, can be resumed in the Mathieu's parameter \emph{q} in order to plots the results in a single plot. \autoref{fig:Cs_q} shows those for a $^{133}$Cs$^+$ beam: in the left there is the transmission and in the right the emittance. Light blue curves are for the tension variation, in which the frequency is set to 4.5 MHz, while the orange ones are for the frequency variation with a tension of 3.45 kV. The pressure inside the HV chamber is 2 Pa. For both the plots the two curves almost overlap, but differ at small values of \emph{q}, where the best performances are reached. The latter are transmission near 70 \% and emittance at about 0.15 $\emittance$ with the chosen \emph{q} equal to 0.06.

\begin{figure}[!tb]
	\centering
	\begin{subfigure}[t]{0.5\textwidth}
		\centering
		\includegraphics[height=5cm]{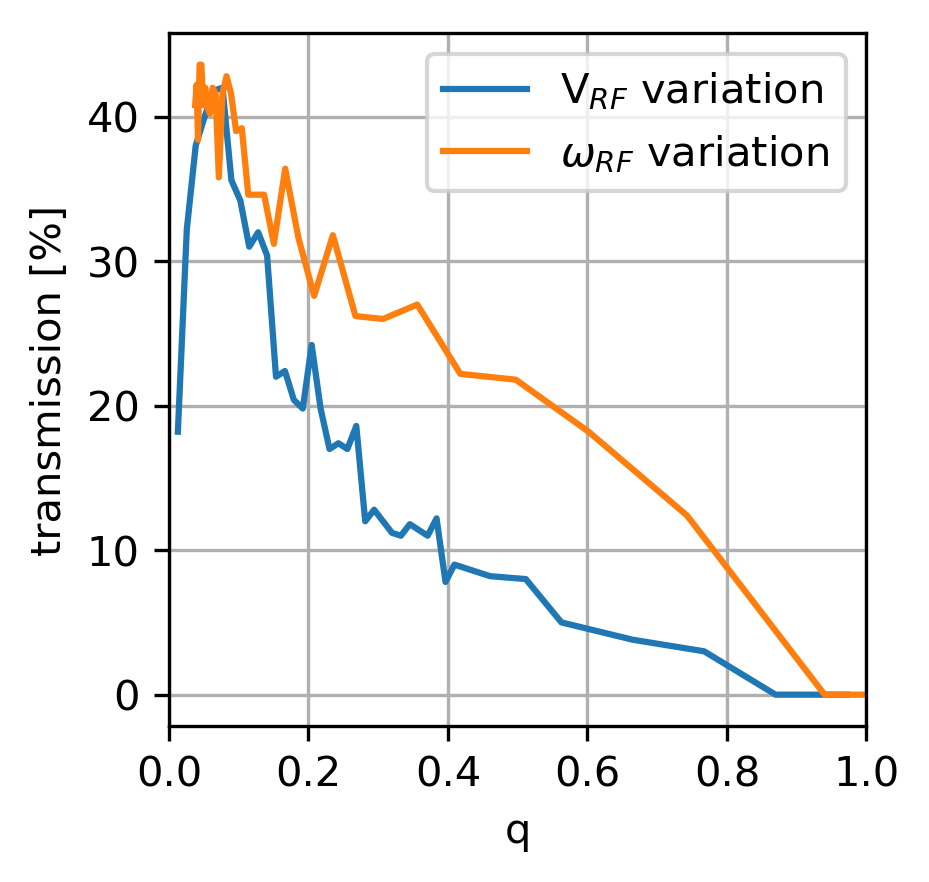}
		\caption{}
		\label{fig:K_transmission_q}
	\end{subfigure}%
	\begin{subfigure}[t]{0.5\textwidth}
		\centering
		\includegraphics[height=5cm]{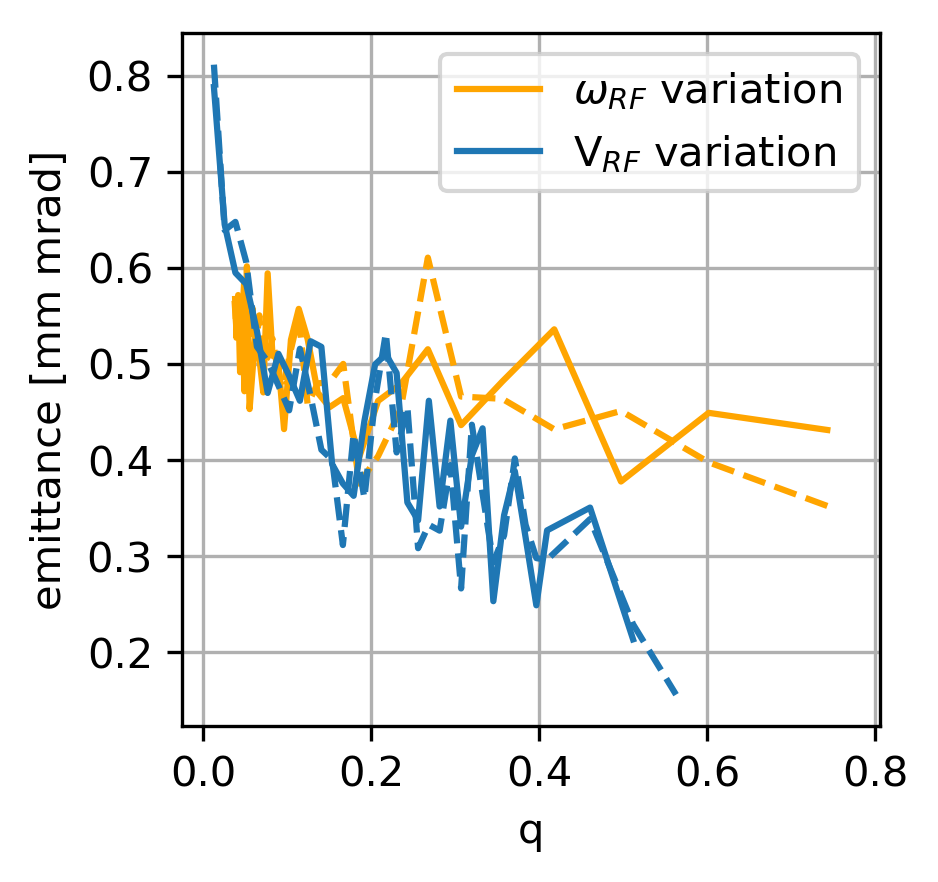}
		\caption{}
		\label{fig:K_emit_q}
	\end{subfigure}
	\caption{Effect of the variation of the Mathieu's parameter \emph{q} in the transmission (left picture) and in the emittance (right picture) for a beam of $^{39}$K$^+$. In the blue curves, \emph{q} is varied changing the polarization potential of the quadrupole, while in the orange curves the polarization frequency changes.}
	\label{fig:K_q}
\end{figure}

Plots in \autoref{fig:K_q} are for a $^{39}$K$^+$ beam. In this case the chosen frequency for the curves generated by the potential variation is 14 MHz, the potential for those given by the frequency variation is set to 1.5 kV and the pressure inside the HV chamber is 1 Pa. For this species light blue and orange curves have a worse overlap along all the range for both parameters, transmission and emittance. Moreover the performances are lower than the $^{133}$Cs$^+$ beam: about 40 \% the best transmission with 0.5 $\emittance$ of emittance polarizing the quadrupole with $q\sim 0.02$.

\begin{figure}[!tb]
	\centering
	\begin{subfigure}[t]{0.5\textwidth}
		\centering
		\includegraphics[height=5cm]{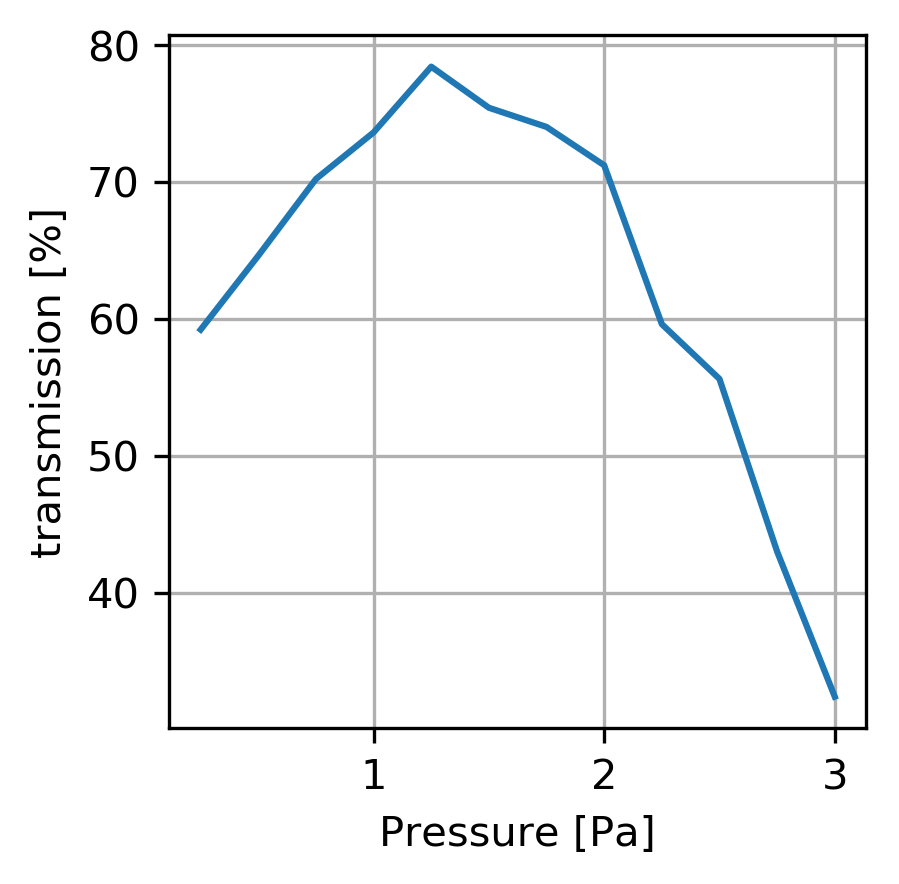}
		\caption{}
		\label{fig:Cs_transmission_P}
	\end{subfigure}%
	\begin{subfigure}[t]{0.5\textwidth}
		\centering
		\includegraphics[height=5cm]{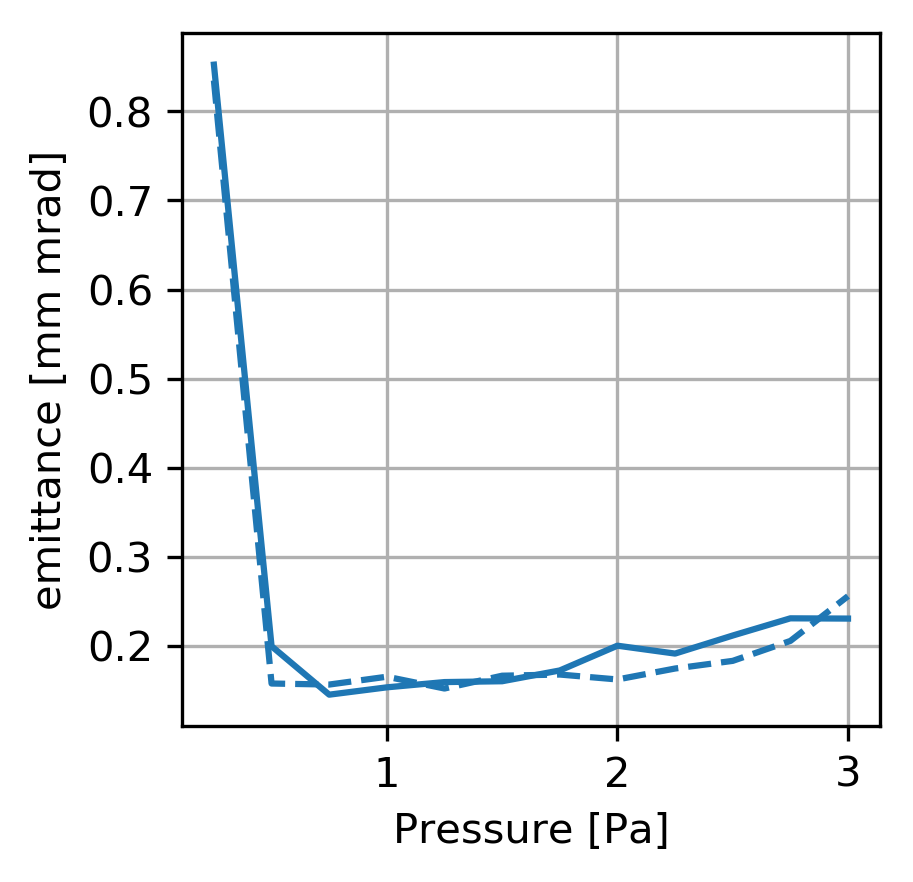}
		\caption{}
		\label{fig:Cs_emit_P}
	\end{subfigure}
	\caption{Variation of the pressure inside the HV chamber, here the ion is $^{133}$Cs$^+$. In the left the transmission and in the right the emittance.}
	\label{fig:Cs_P}
\end{figure}

\autoref{fig:Cs_P} shows what happens to transmission and emittance changing the pressure inside the HV chamber for a beam of $^{133}$Cs$^+$. The best compromise between transmission and emittance is achieved at 1 Pa, although the changes are marginal.

\begin{figure}[!tb]
	\centering
	\begin{subfigure}[t]{0.5\textwidth}
		\centering
		\includegraphics[height=5cm]{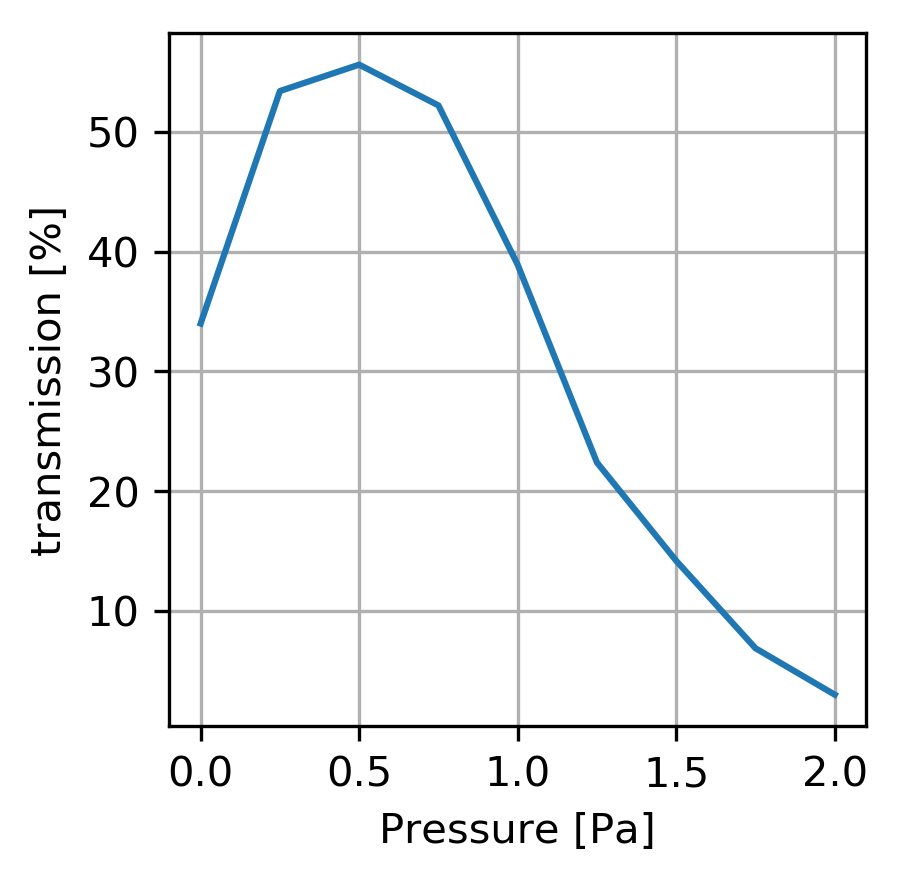}
		\caption{}
		\label{fig:K_transmission_P}
	\end{subfigure}%
	\begin{subfigure}[t]{0.5\textwidth}
		\centering
		\includegraphics[height=5cm]{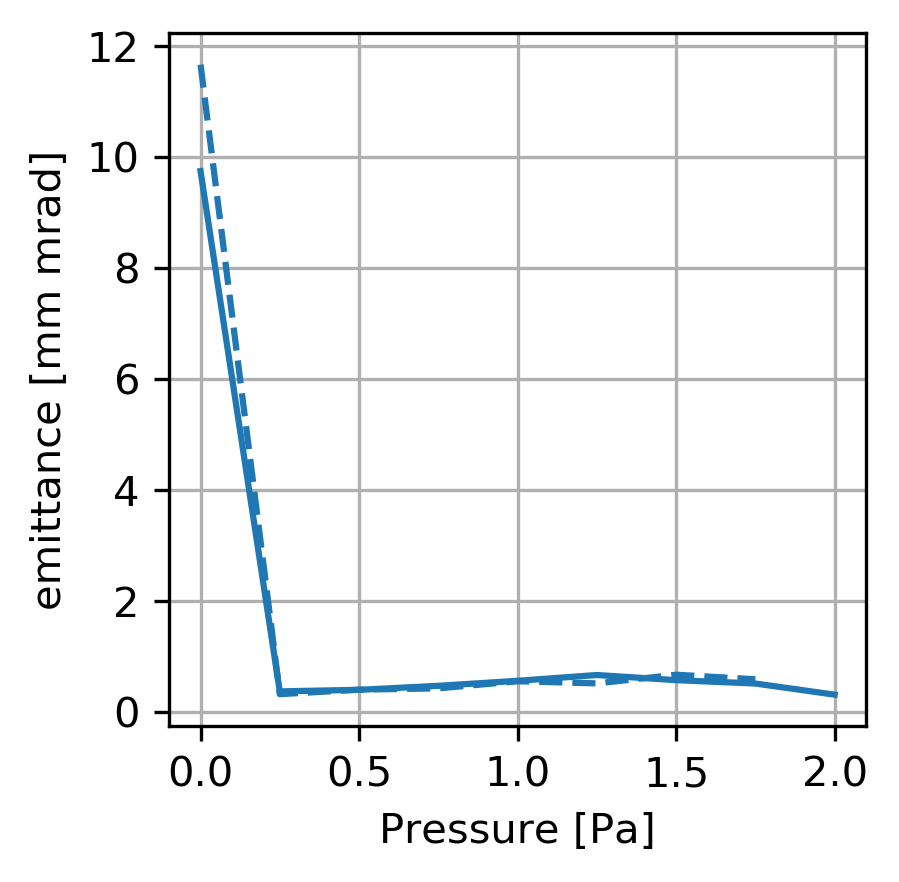}
		\caption{}
		\label{fig:K_emit_P}
	\end{subfigure}
	\caption{Variation of the pressure inside the HV chamber, here the ion is $^{39}$K$^+$. In the left the transmission and in the right the emittance.}
	\label{fig:K_P}
\end{figure}

The same plots are given for a $^{39}$K$^+$ beam in \autoref{fig:K_P}. In this case the effect are significant for the transmission, which surpasses 53 \% from 0.25 Pa to 0.75 Pa. In this case the best performances for both the indexes are reached at 0.25 Pa.

In the following analysis all the simulation parameters except the extraction lens setting are held constant to the best obtained so far. This means for the $^{133}$Cs$^+$ beam: HV chamber voltage sets to 39.74 kV, injection lens electrodes at -0.3, -3.4, -0.3 kV, gas pressure equal to 1 Pa and quadrupoles biased to 1.2 kV @ 7.5 MHz. The $^{39}$K$^+$ beam simulations have: HV chamber voltage equal to 39.74 kV, injection electrodes to -0.5, -2.0, -0.5 kV, pressure of 0.25 Pa and quadrupoles bias 1.5kV @ 16 MHz.



\begin{figure}[!tb]
	\centering
	\begin{subfigure}[t]{0.5\textwidth}
		\centering
		\includegraphics[height=5cm]{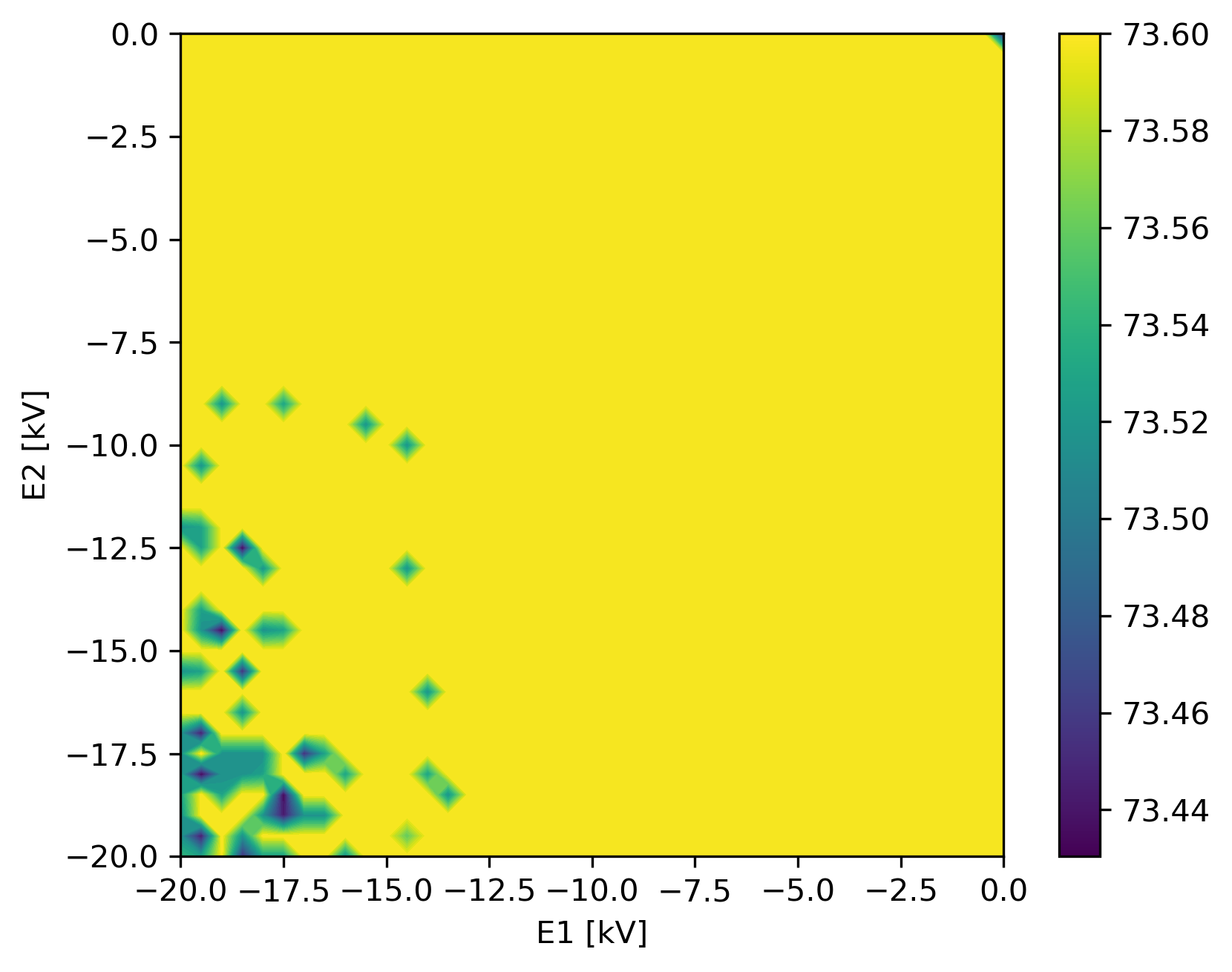}
		\caption{}
		\label{fig:Cs_transmission_extraction}
	\end{subfigure}%
	\begin{subfigure}[t]{0.5\textwidth}
		\centering
		\includegraphics[height=5cm]{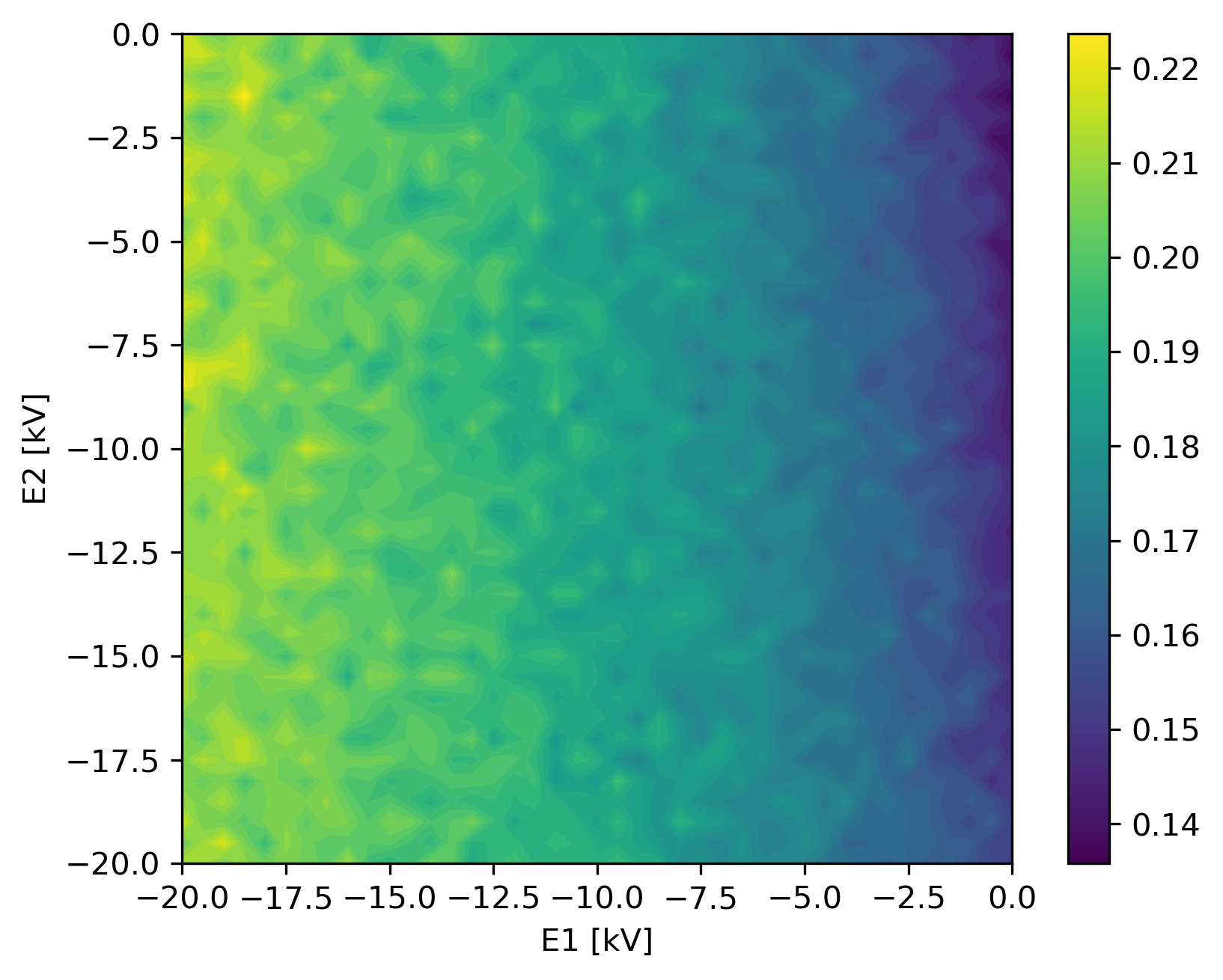}
		\caption{}
		\label{fig:Cs_emit_extraction}
	\end{subfigure}
	\caption{Extraction lens settings effect in the transmission (left) and in the emittance (right). The beam is composed by $^{133}$Cs$^+$.}
	\label{fig:Cs_contour_extraction}
\end{figure}

The results for the $^{133}$Cs$^+$ are in \autoref{fig:Cs_contour_extraction}, transmission in the left and emittance in the right. Here it's possible to see that the extraction lens has not great effect in the transmission, which is stable at 73 \% for all the settings. Concerning the emittance only for the first electrode is effective.

\begin{figure}[!tb]
	\centering
	\begin{subfigure}[t]{0.5\textwidth}
		\centering
		\includegraphics[height=5cm]{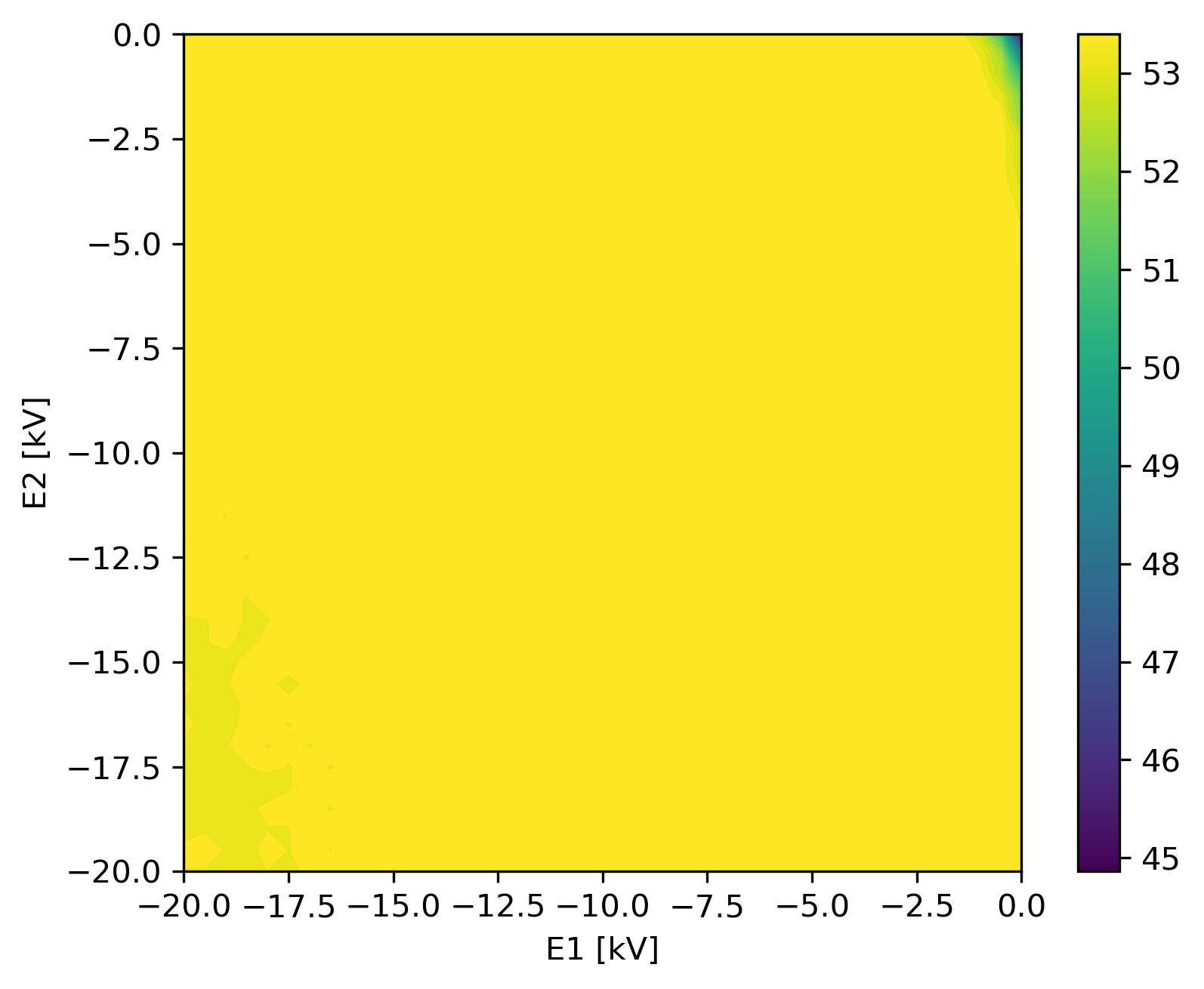}
		\caption{}
		\label{fig:K_transmission_extraction}
	\end{subfigure}%
	\begin{subfigure}[t]{0.5\textwidth}
		\centering
		\includegraphics[height=5cm]{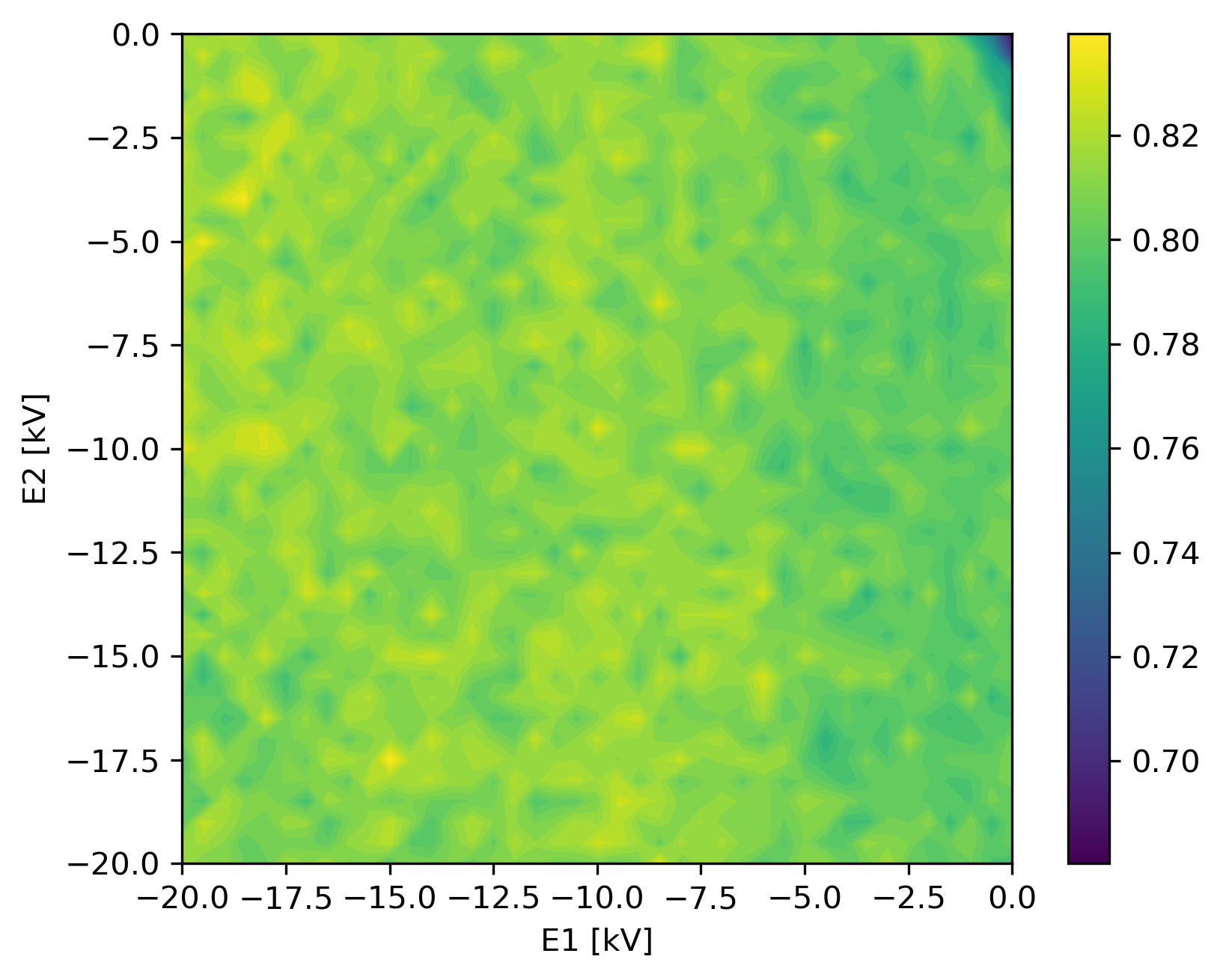}
		\caption{}
		\label{fig:K_emit_extraction}
	\end{subfigure}
	\caption{Extraction lens settings effect in the transmission (left) and in the emittance (right). The beam is composed by $^{39}$K$^+$.}
	\label{fig:K_contour_extraction}
\end{figure}

\autoref{fig:K_contour_extraction} repeats the analyses for a $^{39}$K$^+$ beam. In this case the effects are even less obvious and the only point to avoid is the ($V^{extr}_{E1}$=0,$V^{extr}_{E2}$=0) V since it causes a noticeable drop in transmission.

\begin{figure}[!tb]
	\centering
	\begin{subfigure}[t]{0.5\textwidth}
		\centering
		\includegraphics[height=5cm]{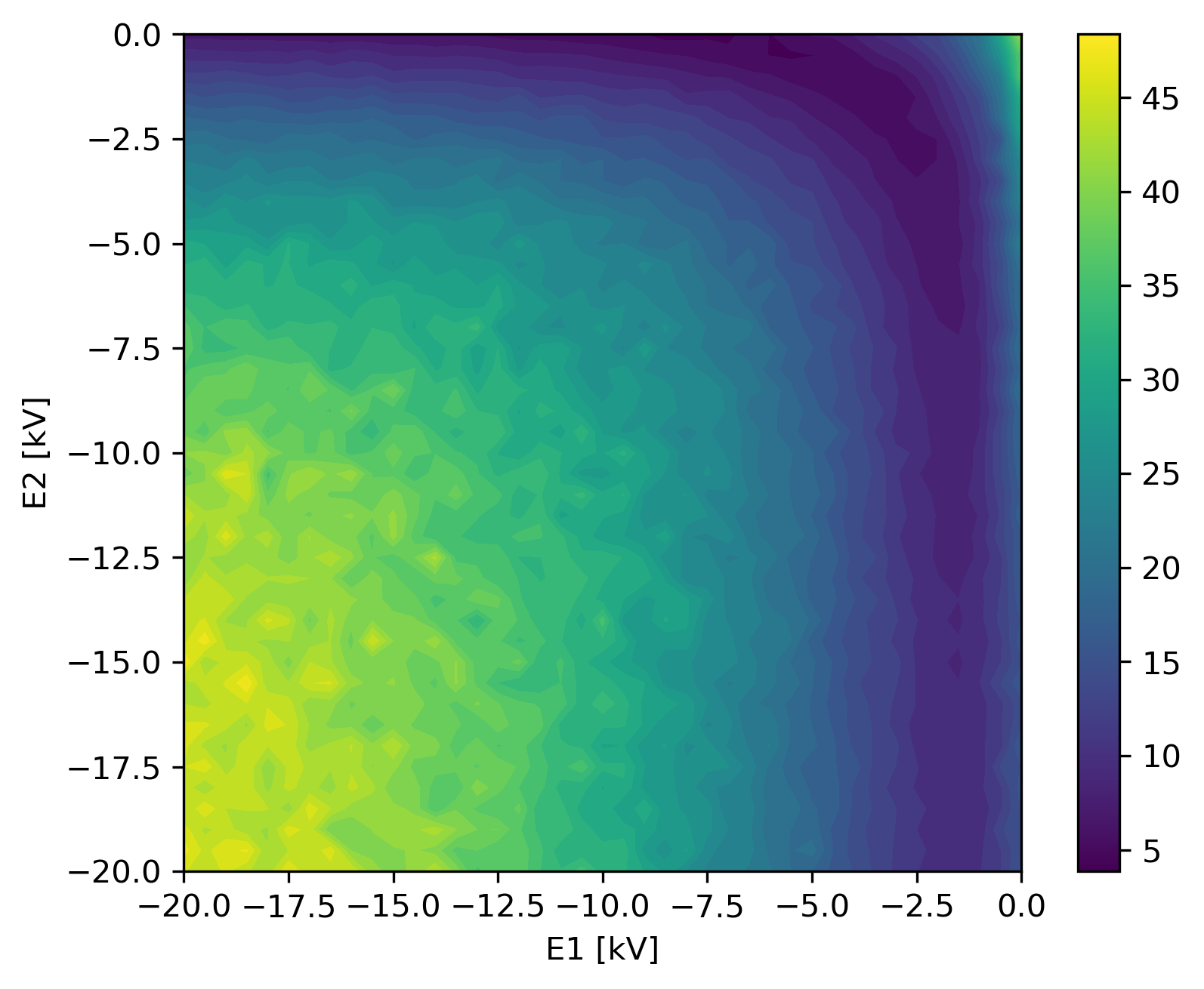}
		\caption{}
		\label{fig:Cs_envelope}
	\end{subfigure}%
	\begin{subfigure}[t]{0.5\textwidth}
		\centering
		\includegraphics[height=5cm]{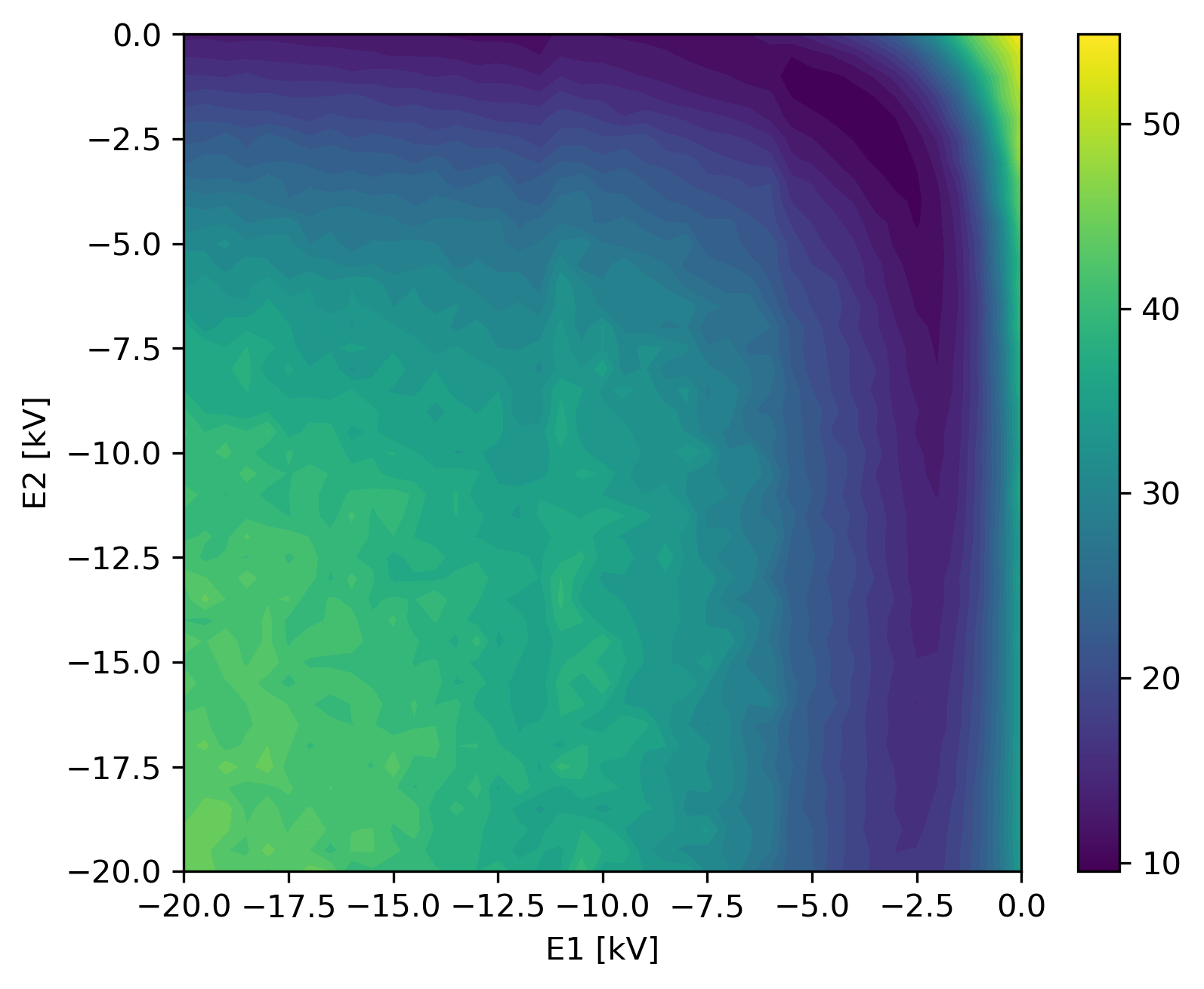}
		\caption{}
		\label{fig:K_envelope}
	\end{subfigure}
	\caption{The envelope of the beam 650 mm after the end of the BC as varies with the potentials of the extraction lens. The left contour is for a caesium beam, the right one is for a Potassium beam.}
	\label{fig:contour_envelope}
\end{figure}

The main purpose of the extraction lens is to allow good beam transport to the first optical element and thus the proper beam width. The \autoref{fig:contour_envelope} shows the envelope 650 mm after the end of the BC for the two species. The smallest width for a $^{133}$Cs$^+$ beam is 4 mm with the electrodes set to the voltage of (-8.5, 0) kV. For the $^{39}$K$^+$ it is 10 mm with the lens set to (-4.5, -1.5) kV.

So far analyses do not take into account the effects of the energy spread, which is one of the most important parameters to perform good selection in a mass separator.

Briefly summarizing what was obtained by varying the settings of the quadrupoles, the energy spread is insensitive to the variation of \emph{q} and oscillates around $\sigma_E^{Cs} = 11$ eV for Caesium and $\sigma_E^{K} = 22$ eV for Potassium.

\begin{figure}[!tb]
	\centering
	\begin{subfigure}[t]{0.5\textwidth}
		\centering
		\includegraphics[height=5cm]{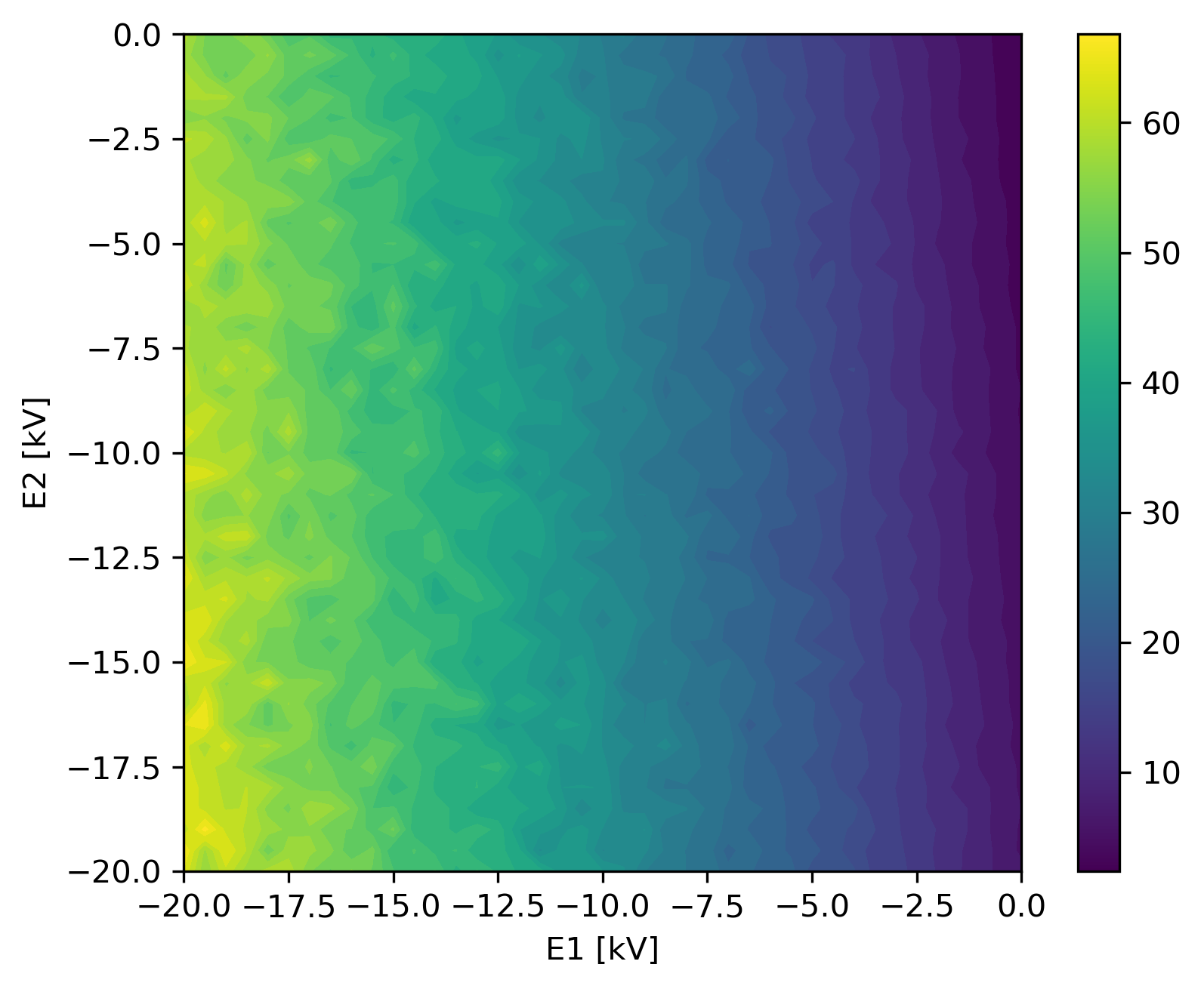}
		\caption{}
		\label{fig:Cs_sigmaE}
	\end{subfigure}%
	\begin{subfigure}[t]{0.5\textwidth}
		\centering
		\includegraphics[height=5cm]{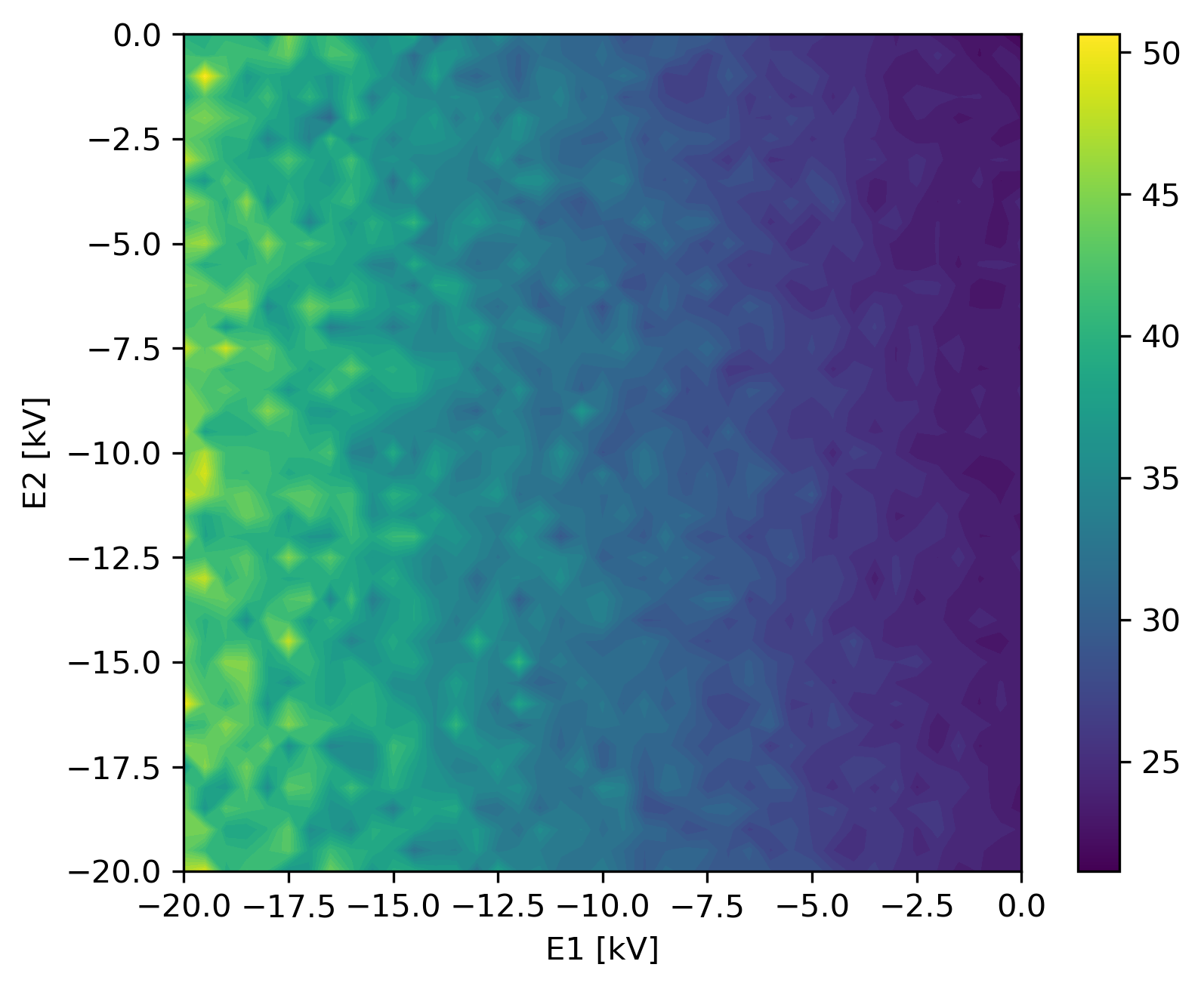}
		\caption{}
		\label{fig:K_sigmaE}
	\end{subfigure}
	\caption{How the extraction lens setting affects the energy spread for a $^{133}$Cs$^+$ beam (left) and a $^{39}$K$^+$ beam (right).}
	\label{fig:contour_sigmaE}
\end{figure}

In the simulations for the extraction lens setup, the energy spread was found to depend mainly on the electric potential of the first electrode. This dependence can be seen in \autoref{fig:contour_sigmaE} for both ion species. The best performance is obtained with the first electrode at the same potential as the HV chamber where it results $\sigma_E^{Cs} = 2.4$ eV for Caesium and $\sigma_E^{K} = 21$ eV for Potassium. Unfortunately, setting the extraction electrodes to 0V does not allow the beam to be focused at the first lens after the BC and further analysis is needed to understand what the allowable limits are.

\begin{figure}[!tb]
	\centering
	\begin{subfigure}[t]{0.5\textwidth}
		\centering
		\includegraphics[height=5cm]{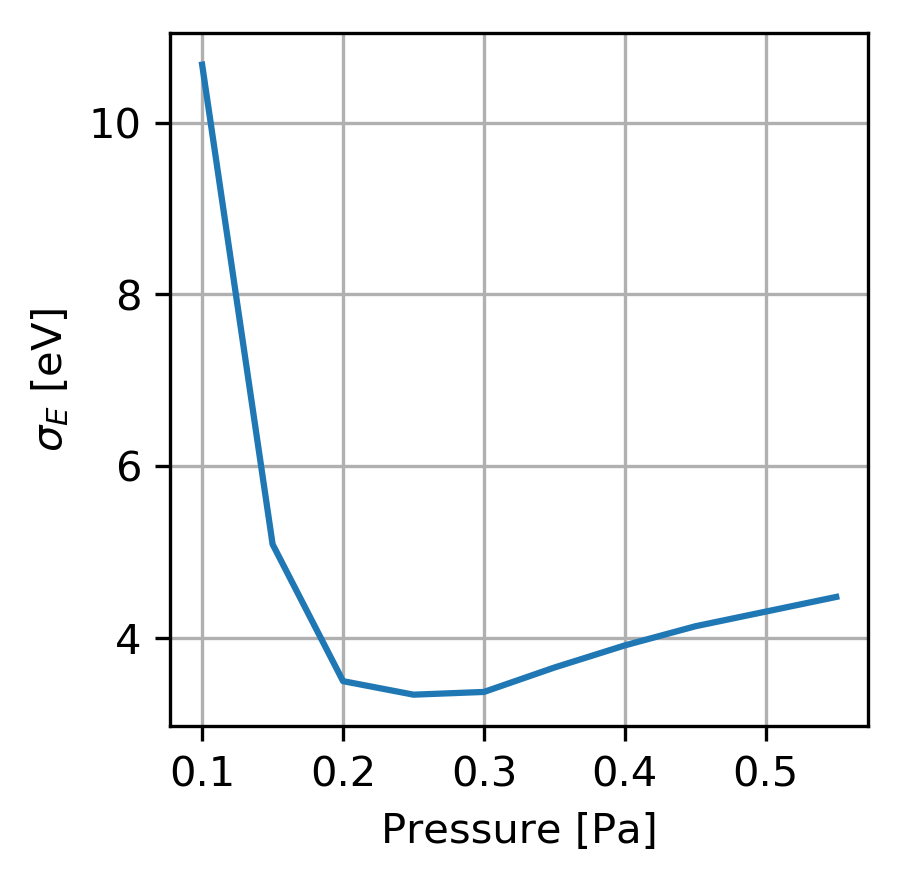}
		\caption{}
		\label{fig:Cs_sigmaE_P}
	\end{subfigure}%
	\begin{subfigure}[t]{0.5\textwidth}
		\centering
		\includegraphics[height=5cm]{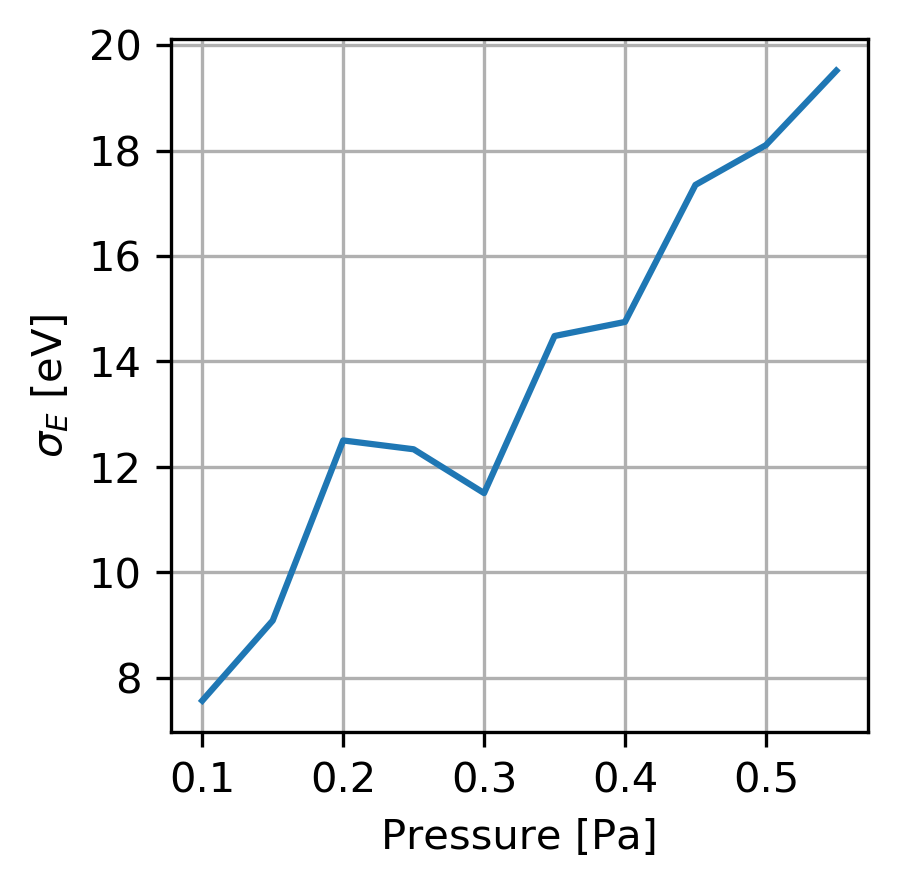}
		\caption{}
		\label{fig:K_sigmaE_K}
	\end{subfigure}
	\caption{The energy spread as a function of the pressure for a $^{133}$Cs$^+$ beam (left) and a $^{39}$K$^+$ beam (right)}
	\label{fig:sigmaE_P}
\end{figure}

The most important effect in the energy spread is given by the pressure. \autoref{fig:sigmaE_P} shows its effect: the lower the pressure, the lower the energy spread. On the contrary, for very low pressures the spread is expected to rise again for the insufficient cooling effect provided by the gas, in the pressure range chosen here this last effect is visible only in the $^{133}$Cs$^+$. In this case, although the results are still improvable, lowering the pressure allows to achieve performance closer to that expected when the BC was designed.

\section{Discussion and improvements} \label{sec:discussion}

The results above say that the BC is very effective in the emittance reducing allowing also a good transmission capacity, especially in the case of heavy ions such as caesium.

In addition to this, some interesting phenomena emerged. First, the injection settings affect the transmission, but not the quality of the extracted beam. This is consistent with the notion that such BC uses a dissipative process to cool the beam.

Another effect can be seen in \autoref{fig:Cs_q} and \autoref{fig:K_q}, they show that using voltage or frequency to vary the Mathieu's parameter \emph{q} is not equivalent, even though theoretically the quadrupoles should act the same way. However the differences are minimal so far.

An interesting fact can be seen in the pressure dependence of the emittance, see \autoref{fig:Cs_emit_P} and \autoref{fig:K_emit_P}: by lowering the pressure it improves to a limit of about 0.5 Pa for the $^{133}$Cs$^+$ and 0.25 Pa for the $^{39}$K$^+$. This is expected since the high gas pressure affects the acceleration phase of the emerging beam, whereas at too much low pressures the cooling effect is reduced.

The thorny issue emerges considering the energy spread, to reduce it the pressure is the most effective parameter, as it results from \autoref{fig:sigmaE_P}. This effect can be explained by the different amount of collisions between ions and gas in the acceleration region: the greater the gas in that region, the greater the number of collisions and thus the energy changes in the emerging beam. The performance shown in \autoref{fig:Cs_contour_extraction} and \autoref{fig:K_contour_extraction} for the extraction lens is also consistent with this result, in fact bringing the potential of the lens closer to that of the HV chamber means moving the acceleration region away from the iris.

Based on this explanation we should try to reduce the presence of gas in the acceleration region and this is the purpose of the proposals made in \autoref{subsec:modified_gas_profile}. To make a comparison, new simulations were performed with the caesium ion: one with the old pressure profile and the other with the new ones while all other parameters are kept the same, namely: $V_{brake}$ = 39.74 kV, injection lens potentials $(-0.3, -3.4, -0.3)$ kV, extraction lens potentials $(-1.5, -2.1)$ kV, quadrupole settings 1.2 kV at 7.5 MHz and gas pressure 0.3 Pa. As a result the energy spread improves from $\sigma_E^{old} = 3.4$ eV to $\sigma_E^{pipe} = 1.0$ eV in the \emph{pipe} case, that is introducing a small pipe before the extraction iris, and to $\sigma_E^{reduced} = 1.5$ eV in the \emph{reduced} case, that is halving the extraction iris dimension. In all of these simulations, also the emittance and the transmission ar not constant passing from 0.33 $\emittance$ in the \emph{old} case to 0.19 $\emittance$ in the new ones and from 59 \% to 54 \%, respectively.

The aim of the BC in the SPES project, like in other facilities, is to prepare the beam for the HMRS and thus, in order to compare the cooling performances, a new parameter can be defined. This should takes into account the dependency of the HRMS resolution from the energy spread and the emittance but should also take into consideration the BC transmission.

In the preparatory studies of the HRMS emerged the dependencies of the resolution as reported in \autoref{eq:risoluzione_dipendenza} where $R$ is the resolution function, $\varepsilon$ the emittance, $\sigma_E$ the energy spread and the two constants, $k_\varepsilon$ and $k_{\sigma_E}$, depend on the beam and on the separator features \cite{comunian_risoluzione_HRMS}; in the SPES case, their values are 3.6$\cdot 10^{-5}$ and 10$\cdot 10^{-5}$ respectively.

\begin{eqnarray}
	R(\varepsilon, \sigma_E) = \sqrt{k_\varepsilon \cdot \varepsilon^2 + k_{\sigma_E} \cdot \sigma_E^2 } \label{eq:risoluzione_dipendenza}
\end{eqnarray}

Starting from the dependencies in \autoref{eq:risoluzione_dipendenza}, one can define the new parameter as in \autoref{eq:new_parameter}, where $Tr$ is the BC transmission.

\begin{eqnarray}
	C = \frac{Tr}{R(\varepsilon, \sigma_E)} \label{eq:new_parameter}
\end{eqnarray}

Following this definition the expected performances for the presented BC with the most significant settings found so far, are shown in the \autoref{tab:coolIndex}.

\begin{table}[!b]
	\centering
	\caption{The cooling index, as defined in \autoref{eq:new_parameter}, for the most significant settings.}
\label{tab:coolIndex}
	\begin{tabular}{l|c|c|c|c}
		\rowcolor{black!90} \textcolor{white}{Settings} & \textcolor{white}{$\sigma_E$} & \textcolor{white}{$\varepsilon^{rms}$ } & \textcolor{white}{$Tr$} & \textcolor{white}{Cooling Index}   \\
		Cs, q=0.06, P=0.3 Pa & 3.4 & 0.33 & 59.3 & 1757 \\ \hline
		Cs, q=0.06, P=0.3 Pa, \emph{pipe} & 1.0 & 0.19 & 53.7 & 5403 \\ \hline
		Cs, q=0.06, P=0.3 Pa, \emph{reduced} & 1.5 & 0.19 & 54.3 & 3570 \\ \hline
		K, q=0.06, P=0.3 Pa & 11.6 & 0.69 & 65.1 & 563 \\ \hline
	\end{tabular}
\end{table}



\section{Conclusions} \label{sec:conclusions}

In the first part of the present document (\autoref{sec:intro} and \autoref{sec:the_cooler}) the new BC, designed and built at LPC, Caen, France, for the new SPES facility at the LNL, Padua, Italy has been detailed. Such a device exploits the dissipative ions-gas interaction to cool down the beam with the main purpose to get the best separation of heavy ions in the following HRMS system.

The chosen gas is $^4$He which is confined in a HV chamber where the cooling process takes place. Such a chamber is 730 mm long and it is characterized by the two irises whose diameters have been kept as small as possible in order to achieve the best gas confinement. Inside that chamber there is a row of 18 RF quadrupoles that have the role to confine ions during their interaction with the gas.

HV chamber has an electric potential close to the kinetic energy of the beam, so that it is slowed down to few hundreds eV. The beam exiting the BC is accelerated again close to the previous energy by the same potential. Before and after this chamber, during beam acceleration, there are the injection and the extraction lenses consisting of three and two electrodes respectively. The first is intended to inject as much of the beam as possible inside the HV chamber, the latter is to provide a focusing effect that allow the beam to reach the first optical element after the BC.

In the second part of this document (\autoref{sec:gas_profile}) the gas confinement performance has been estimated thanks to the MolFlow+ code. The pressure profile is useful to have consistent results in the subsequent beam dynamic simulations. Since the gas confinement is crucial to the extracted beam quality, two improvements are also presented.

In the last part (\autoref{sec:simion_simulation} and \autoref{sec:discussion}) beam dynamic simulations are presented and discussed. They help to frame the performance provided by the BC and, by varying all the available parameters that configure the device, to find those that are expected to perform better.

Simulations show that the BC can provide a maximum transmission above the 70 \% for heavy ions and 55 \% for light ions for a starting beam with 10 $\emittance$ emittance. Moreover the emerging beam have an emittance lower to 0.2 $\emittance$ for the $^{133}$Cs$^+$ and 0.5 $\emittance$ for the $^{39}$K$^+$. Although the good performance in the transverse dimension, the same settings imply quite high value of energy spread ($\sigma_E^{Cs} > 10$ eV and $\sigma_E^{K} > 20$ eV). To reduce it the most effective parameter resulted the pressure: if it is lowered to very low values, somewhere close to 0.3 Pa for both heavy and light ions, energy spread will improve to $\sigma_E^{Cs} = 3.4$ eV and $\sigma_E^{K} = 11$ eV.

This performance can be further improved by varying the geometry of the extraction iris to limit the gas loss. The \autoref{subsec:modified_gas_profile} presents some ideas studied in the next section with a $^{133}$Cs$^+$ beam. With this addition, the energy diffusion performance is improved by three to two times.


%
%

\bibliographystyle{JHEP}
\bibliography{bibliografia_Ruzzon.bib}

\end{document}